\documentclass[journals,a4paper]{IEEEtran}
\ifCLASSINFOpdf
\else
\usepackage[dvips]{graphicx}
\fi
\usepackage[cmex10]{amsmath}
\usepackage{upgreek}
\usepackage{booktabs}
\usepackage{epsfig}
\usepackage{latexsym}
\usepackage{multirow}
\usepackage{stfloats}
\usepackage{epstopdf}
\usepackage{color}  
\usepackage{tabularx} 
\usepackage{amssymb}
\usepackage{enumerate}
\graphicspath{{./Figures/}}
\usepackage{color}
\usepackage{bbm}
\usepackage{bm}
\usepackage{pifont}
\usepackage{cite}
\usepackage{balance}
\usepackage{threeparttable}
\usepackage{mathrsfs}
\usepackage{verbatim}
\usepackage{dsfont}
\usepackage{verbatim}
\usepackage{tikz}
\usepackage{diagbox}
\usepackage{caption}
\usepackage[framemethod=tikz]{mdframed}
\usepackage{multicol}
\usepackage{environ}
\usepackage{tikz}

\def\dif{\mathop{}\hphantom{\mskip-\thinmuskip}\mathrm{d}}%
\let\daccent\d
\let\d\relax
\newcommand\d{\ifmmode\dif\else\expandafter\daccent\fi}

\begin{document}

\title{
A Universal Attenuation Model of Terahertz Wave in Space-Air-Ground Channel Medium}
\author{\IEEEauthorblockN{$\textrm{Zhirong Yang}, \textrm{Weijun Gao}, \textit{Graduate Student Member, IEEE}$, and $\textrm{Chong Han}, \textit{Member, IEEE}$
	}

\thanks{
Zhirong~Yang and Weijun~Gao are with the Terahertz Wireless Communications (TWC) Laboratory, Shanghai Jiao Tong University, Shanghai, China (Email:~\{zhirong.yang, gaoweijun\}@sjtu.edu.cn). 

Chong~Han is with the Terahertz Wireless Communications (TWC) Laboratory, Department of Electronic Engineering and Cooperative Medianet Innovation Center (CMIC), Shanghai Jiao Tong University, Shanghai, China (Email:~chong.han@sjtu.edu.cn). }
}

\markboth{IEEE Open Journal of the Communications Society, submitted in July 2023}{}
\maketitle
\thispagestyle{empty}

\begin{abstract}
Providing continuous bandwidth over several tens of GHz, the Terahertz (THz) band (0.1-10 THz) supports space-air-ground integrated network (SAGIN) in 6G and beyond wireless networks.
However, it is still mystery how THz waves interact with the channel medium in SAGIN.
In this paper, a universal space-air-ground attenuation model is proposed for THz waves, which incorporates the attenuation effects induced by particles including condensed particles, molecules, and free electrons.
The proposed model is developed from the insight into the attenuation effects, namely, the physical picture that attenuation is the result of collision between photons that are the essence of THz waves and particles in the environment.
Based on the attenuation model, the propagation loss of THz waves in the atmosphere and the outer space are numerically assessed.
The results indicate that the attenuation effects except free space loss are all negligible at the altitude higher than 50~km while they need to be considered in the atmosphere lower than 50~km.
Furthermore, the capacities of THz SAGIN are evaluated in space-ground, space-sea, ground-sea, and sea-sea scenarios, respectively.
\end{abstract}

\IEEEpeerreviewmaketitle

\section{Introduction}\label{sec:Intro}

Space-air-ground integrated network (SAGIN), as an emerging communication scenario to provide global coverage for 6G wireless networks, puts a stringent requirement for the high data rates of wireless communications~\cite{8368236,9222519,9520380,9583591}.
The Terahertz (THz) band (0.1-10 THz) has been envisioned as one of the key solutions to support the Terabit-per-second (Tbps) data traffic in future 6G and beyond wireless networks.
Thanks to the abundant bandwidth of the THz band, THz SAGIN has the potential to realize ultra-high-speed wireless data transmission ubiquitously and globally, such that the emerging applications and technologies, including ubiquitous artificial intelligence (AI), industrial internet of things (IIoT), digital twins (DT) and etc., can be supported~\cite{9846955,9858592,9931961}.

Establishing THz space-air-ground links needs careful analysis of the propagation loss caused by various attenuation effects.
In addition to free-space path loss (FSPL), THz waves experience absorption and scattering effects induced by the particles encountered during propagation.
On one hand, some effects can cause severer attenuation for THz waves compared with the lower frequencies.
First, the FSPL of THz waves is higher due to the smaller effective area of THz antennas.
Second, water vapor can significantly attenuate THz waves through absorption, which is negligible for the lower frequencies.
Third, particles like cloud droplets can induce more attenuation for the shorter wavelength and weaker diffraction of THz waves.
On the other hand, compared to the lower frequencies, THz waves suffer from fewer attenuation in plasma, such as the ionosphere and solar wind~\cite{9541155}.
Thus, it is necessary to precisely evaluate the propagation loss of THz space-air-ground links such that the requirement on link gain and the feasibility of THz SAGIN can be clarified.

In the literature, existing attenuation models for the THz wave propagation characterize only a certain portions of attenuation effects.
For instance, the absorption loss of THz waves in plasma is modeled in~\cite{9541155}, which is utilized to assess the capacity of inter-satellite links.
A conventional THz aerial channel model considers both the molecular absorption effect and the loss caused by cloud~\cite{9352550}. 
Moreover, in the recommendation of international telecommunication union (ITU), there are three attenuation models, namely ITU-R P.676-13 for molecular absorption loss in the atmosphere~\cite{ITU676}, ITU-R P.840-8 for the attenuation caused by cloud and fog~\cite{ITU840}, and ITU-R P.838-3 for the attenuation induced by rain~\cite{ITU838}.
These models are empirical models and only considers certain atmospheric effect, while a universal model that incorporates all kinds of effect in SAGIN is still missing.
The remaining problems in ITU standard models include:
\begin{itemize}
    \item ITU-R P.676-13 does not show a perfect accordance with a THz channel model for nano-networks which uses the data from high resolution transmission molecular absorption database (HITRAN)~\cite{ITU676,5995306}. 
    \item ITU-R P.838-3 does not incorporate the impact of different rain spectra, i.e., the distributions of the diameter of raindrops. However, as pointed out by~\cite{8826596}, different rain spectra can lead to different rain attenuation of THz wave propagation.
    \item ITU-R P.840-8 does not consider the altitude dependency of cloud or fog attenuation, which limits its application in THz SAGIN with varying altitudes.
\end{itemize}
As listed in TABLE~\ref{tab:saga}, the attenuation effects that THz waves encounter in the atmosphere and space include Mie absorption and scattering, Rayleigh absorption and scattering, molecular absorption and scattering, Coulomb absorption, and Thomson scattering~\cite{9567386,10138913}.
There exists no universal attenuation models considering all these attenuation effects to characterize THz wave propagation from space, air to ground or vice versa~\cite{9444237}.

To address the aforementioned issues, this paper proposes a THz space-air-ground attenuation model based on the physical picture that THz waves are photons in essence~\cite{doi:10.1098/rspa.1927.0039}.
In the physical picture, the interaction between THz waves and particles in medium are the collision between photons and various particles in medium, so that various attenuation effects are incorporated under the same scheme.
Furthermore, we analyze the propagation loss of THz waves in different layers of the atmosphere, including the troposphere with altitudes ranging from 0 to 10 km, the stratosphere from 10 to 50 km, the ionosphere from 50 to 1000 km, and the outer space above 1000 km. 
Moreover, we compare the proposed model and ITU models interest of rain attenuation and molecular absorption loss.
In addition, we assess the capacity of THz SAGIN in space-ground and sea-sea scenarios.

\begin{table*}[htbp]
\centering
\caption{Attenuation effects on THz waves in space-air-ground channel medium}
\label{tab:saga}
\begin{tabular}{|c|cc|cc|cc|cc|c|}
\hline
Particles & \multicolumn{2}{c|}{\begin{tabular}[c]{@{}c@{}}Condensed particles\\ $d\approx\lambda$\end{tabular}} & \multicolumn{2}{c|}{\begin{tabular}[c]{@{}c@{}}Condensed particles\\ $d\ll\lambda$\end{tabular}} & \multicolumn{2}{c|}{Molecules} & \multicolumn{2}{c|}{Free electrons} &  \\ \hline
\begin{tabular}[c]{@{}c@{}}Attenuation\\ effects\end{tabular} & \multicolumn{1}{c|}{\begin{tabular}[c]{@{}c@{}}Mie\\ absorption\end{tabular}} & \begin{tabular}[c]{@{}c@{}}Mie\\ scattering\end{tabular} & \multicolumn{1}{c|}{\begin{tabular}[c]{@{}c@{}}Rayleigh\\ absorption\end{tabular}} & \begin{tabular}[c]{@{}c@{}}Rayleigh\\ scattering\end{tabular} & \multicolumn{1}{c|}{\begin{tabular}[c]{@{}c@{}}Molecular\\ absorption\end{tabular}} & \begin{tabular}[c]{@{}c@{}}Rayleigh\\ scattering\end{tabular} & \multicolumn{1}{c|}{\begin{tabular}[c]{@{}c@{}}Coulomb\\ absorption\end{tabular}} & \begin{tabular}[c]{@{}c@{}}Thomson\\ scattering\end{tabular} & \begin{tabular}[c]{@{}c@{}}Free-space\\path loss\end{tabular} \\ \hline
Troposphere & \multicolumn{1}{c|}{\checkmark} & \checkmark & \multicolumn{1}{c|}{\checkmark} & \checkmark & \multicolumn{1}{c|}{\checkmark} & \checkmark & \multicolumn{1}{c|}{\ding{53}} & \ding{53} & \checkmark \\ \hline
Stratosphere & \multicolumn{1}{c|}{\ding{53}} & \ding{53} & \multicolumn{1}{c|}{\ding{53}} & \ding{53} & \multicolumn{1}{c|}{\checkmark} & \checkmark & \multicolumn{1}{c|}{\ding{53}} & \ding{53} & \checkmark \\ \hline
Ionosphere & \multicolumn{1}{c|}{\ding{53}} & \ding{53} & \multicolumn{1}{c|}{\ding{53}} & \ding{53} & \multicolumn{1}{c|}{\ding{53}} & \ding{53} & \multicolumn{1}{c|}{\checkmark} & \checkmark & \checkmark \\ \hline
Outer space & \multicolumn{1}{c|}{\ding{53}} & \ding{53} & \multicolumn{1}{c|}{\ding{53}} & \ding{53} & \multicolumn{1}{c|}{\ding{53}} & \ding{53} & \multicolumn{1}{c|}{\checkmark} & \checkmark & \checkmark \\ \hline
\end{tabular}
\end{table*}
\begin{figure*}[!h]
    \centering
    \includegraphics[width=0.9\textwidth]{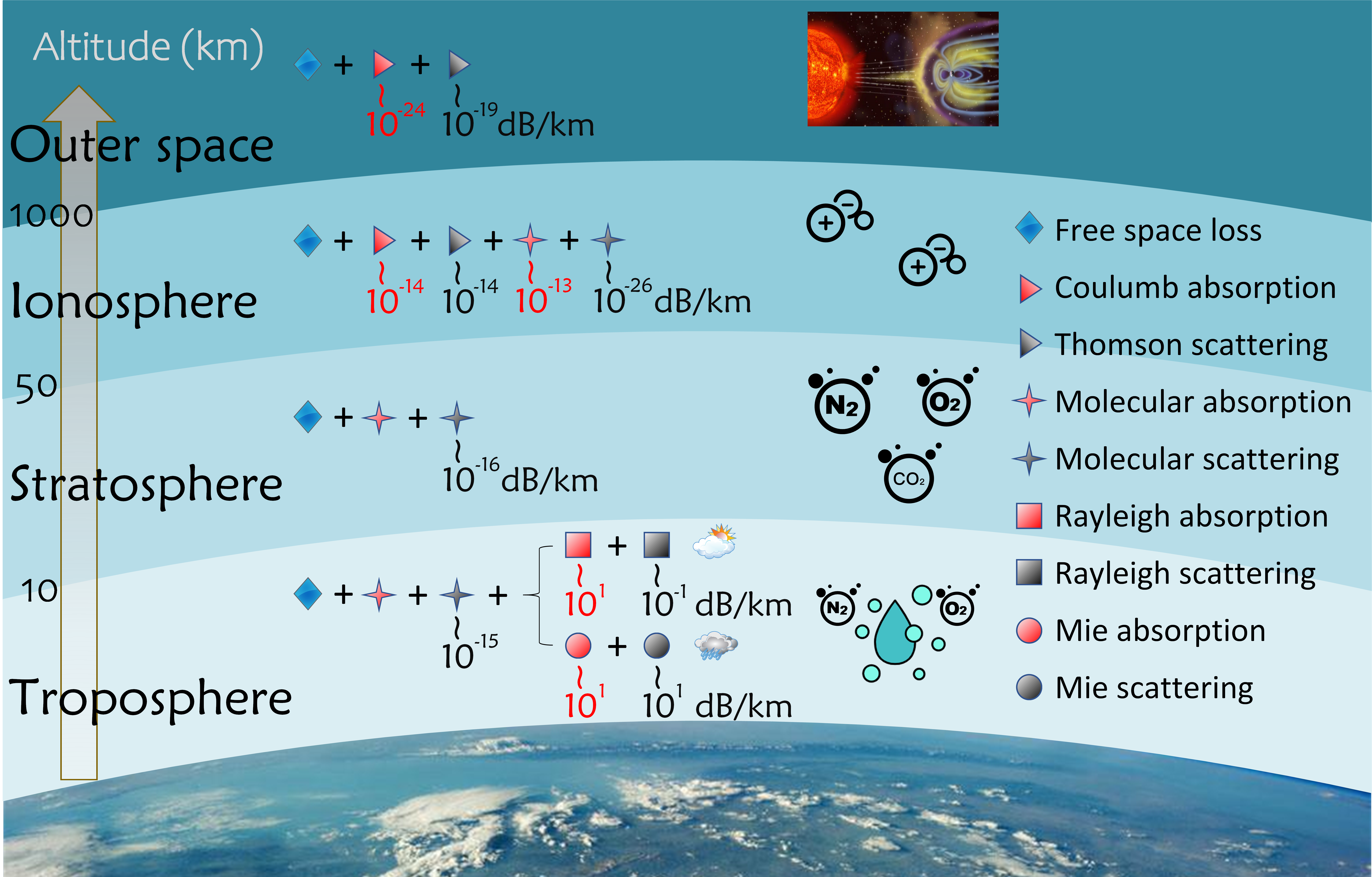}
    \captionsetup{font=footnotesize}
    \caption{THz wave propagation in the atmosphere and the outer space.}
    \label{fig_Tpropa}
\end{figure*}

The contributions of this work are summarized as follows:
\begin{itemize}
    \item We propose a universal attenuation model based on the physical picture of collision for THz ground-air-space communications, which incorporates the common attenuation effects on THz waves in the atmosphere, namely, Mie absorption and scattering, Rayleigh absorption and scattering, molecular absorption and scattering, Coulomb absorption, and Thomson scattering.
    \item We model the spatial distribution of particles in the atmosphere and evaluate the propagation loss in various effects in different layers of the atmosphere and the outer space, as shown in TABLE~\ref{tab:saga} and Fig.~\ref{fig_Tpropa}.
    \item We compare and contrast the proposed model and the ITU models. For rain attenuation, the proposed model can incorporate the environment-dependent rain spectra, while the empirical model in ITU-R P.838-3 does not consider the different rain spectra in various scenarios.
    For molecular absorption, the proposed model fits well with ITU-R P.676-13 and has higher interpretability and can incorporate more attenuation effects except molecular absorption.
    \item We compute the capacity of THz SAGIN in space-ground, space-sea, ground-sea, and sea-sea scenarios. 
    The results demonstrate that the capacity of the link in the troposphere is the bottleneck of THz SAGIN in space-sea and space-ground scenarios. 
    Moreover, communication links cannot be established with less than two relay nodes in ground-sea scenario and the band in 151.5-164~GHz can achieve the highest capacity in the sea-sea scenario in THz SAGIN.
\end{itemize}

The rest of this paper is presented as follows. 
The attenuation model based on the physical picture of collision is elaborated in Sec.~\ref{sec:AttMod}. 
The propagation loss of different layers of the atmosphere is assessed in Sec.~\ref{sect:TPL}.
In Sec.~\ref{sec:comitu}, the proposed model is compared with the ITU models.
In Sec.~\ref{sec:comitu}, the capacity of THz SAGIN is assessed.
Finally, Sec.~\ref{sec:conclusion} concludes the paper.

\textit{Notation:} 
Operator $\dif$ represents the differential operation.
$d$ stands for the diameter of particle.
$\operatorname{Re}(\cdot)$ returns the real part of a complex number.
$(\cdot)'$ refers to the derivative. 
$\exp(\cdot)$ defines the exponential function.
$j_m(\cdot)$ and $h_m(\cdot)$ denote the spherical functions of Bessel and Hankel of the first kind of order $m$, respectively.


\section{Space-Air-Ground Attenuation Model}\label{sec:AttMod}
During the propagation of THz waves in the atmosphere and space, THz waves suffer from different attenuation effects in addition to FSPL as illustrated in Fig.~\ref{fig_Tpropa}.
Various components in the channel medium from the troposphere to the outer space, such as raindrops, fog drops, atmospheric molecules, free electrons and etc., induce multifarious attenuation effects on Thz waves.
A comprehensive scheme is thus required to model the attenuation during THz wave propagation in space-air-ground channel medium. 
In this section, the physical picture of THz wave propagation is analyzed to comprehensively incorporate various attenuation effects, including Mie absorption and scattering, Rayleigh absorption and scattering, molecular absorption and scattering, Coulomb absorption, and Thomson scattering.
The total propagation loss model in space-air-ground channel is developed based on the physical picture.


\subsection{Physical Picture of Attenuation}

Physically from the micro-scale perspective, the attenuation effects are essentially the collisions between photons that are the essence of THz waves and various particles in the environment.
We consider a cylindrical area along the propagation path with cross section $A$ and length $\dif s$. 
We assume that the particles in the area are identical particles of which the chemical composition and the size are all the same, and the cross section of a particle in this area is denoted by $\sigma$. 
Therefore, the probability that a collision occurs can be expressed by,
\begin{equation}
    \textrm{Pr}=\frac{\sigma N}{A}
    \label{eq:Pr}
\end{equation}
where $N$ denotes the number of particles in this area. Let $n$ denote the number density of the particles, i.e., the number of particles per unit volume, as $n=N/A\dif s$. 
When a collision happens, the incident photon can be absorbed or scattered by the particle.
Therefore, we further divide the cross section into two cross sections $\sigma_{A}$ and $\sigma_{S}$, as absorption cross section and scattering cross section of the particle, i.e., $\sigma=\sigma_{A}+\sigma_{S}$. 
Therefore, the collision probability can be expressed as,
\begin{equation}
\operatorname{Pr}=n\left(\sigma_{A}+\sigma_{S}\right)\dif s.
\end{equation}
The cross sections of a particle are parameters in units of area, which are utilized for the characterization of the probability that the corresponding effect occurs. 
Unless otherwise noted, the parameters of the formulas in this paper are all in the International Systems of Units.
Since THz waves are a group of photons, the number density of photons is proportional to the intensity of THz waves, denoted by $I$.
Hence, the variation of the intensity of THz waves, denoted by $\dif I$, is proportional to the number density of the photons that collide with particles and given as,
\begin{equation}\label{eq:dI}
    \dif I=-I\cdot\operatorname{Pr}.
\end{equation}
Then, the intensity of the outgoing THz waves is derived as,
\begin{equation}\label{eq:ILS}
    I=I_0\exp\left[{-\left(\sigma_{A}+\sigma_{S}\right)\cdot\int_{s_0}^{s_1}n\left(s\right)\dif s}\right],
\end{equation}
where $I_0$ refers to the intensity of incident waves, $s_{0}$ and $s_{1}$ represent the origin and the end of the propagation in the environment, respectively. 

Since there are multiple types of particles with different composition or sizes in the atmosphere, such as raindrops, cloud droplets, atmospheric molecules, free electrons, and etc., the equation~\eqref{eq:ILS} is rewritten as,
\begin{equation}\label{eq:ILM}
    I=I_0\exp\left[{-\sum_{\varsigma}\left(\sigma_{A|\varsigma}+\sigma_{S|\varsigma}\right)\cdot\int_{s_0}^{s_1}n_{\varsigma}\left(s\right)\dif s}\right],
\end{equation}
where the subscript $\varsigma$ denotes the type of particles.
It is worth noting that only identical particles can be classified into the same type, while the particles with the same chemical composition and various diameters are of different types since the cross sections vary with the diameters of particles.
Then, the loss caused by collisions with particles, called collisional loss, is defined as,
\begin{equation}\label{eq:LC}
    \begin{aligned}
        L_{C}=&-10\log_{10}\frac{I_{}}{I_0}\\        =&10\log_{10}e\cdot\sum_{\varsigma}\left(\sigma_{A|\varsigma}+\sigma_{S|\varsigma}\right)\cdot\int_{s_0}^{s_1}n_{\varsigma}\left(s\right)\dif s,
    \end{aligned}
\end{equation}
where $L_C$ is in the unit of dB.


We then introduce the cross sections of different particles in the atmosphere, including a condensed particle composed of numerous molecules, a single molecule, and a free electron, which correspond to various attenuation effects and expressions.
Specifically, the relationship is: 
(1). the condensed particles with diameter~$d\approx\lambda$ correspond to Mie absorption and scattering,
(2). the condensed particles with diameter~$d\ll\lambda$ correspond to Rayleigh absorption and scattering, 
(3). molecules correspond to molecular absorption and scattering, 
(4). free electrons correspond to Coulomb absorption and Thomson scattering, 
where $\lambda$ denotes the wavelength of THz waves.
Note that we do not consider the case $d\gg\lambda$ since the diameter $d$ of the particles in the atmosphere are generally not much larger than the wavelength $\lambda$ of THz waves~\cite{10023042}. 

\subsection{Mie Absorption and Scattering}

For condensed particles, which are composed of numerous molecules and generally in condensed states such as liquid and solid, the absorption cross section $\sigma_{A|M}$, named after Mie absorption cross section, is given as~\cite{9783527618156},
\begin{subequations}
    \begin{align}
        \sigma_{A|M}=&\frac{\pi d^2}{2\xi^2}\sum_{m=1}^\infty\left(2m+1\right)\cdot\notag
        \\&\left[\operatorname{Re}\left(a_m+b_m\right)        -|a_m|^2-|b_m|^2\right]\\
        \approx&\frac{\pi d^2}{2\xi^2}\sum_{m=1}^{m_{\operatorname{max}}}\left(2m+1\right)\cdot\notag
        \\&\left[\operatorname{Re}\left(a_m+b_m\right)-|a_m|^2-|b_m|^2\right],\label{eq:AM}
    \end{align}
\end{subequations}
with $\xi=\pi d/\lambda$ and,
\begin{subequations}
    \begin{align}
        a_m=&\frac{q^2j_m(q\xi)[\xi j_m(\xi)]'-\mu_1j_m(\xi)[q\xi j_m(q\xi)]'}{q^2j_m(q\xi)[\xi h_m(\xi)]'-\mu_1h_m(\xi)[q\xi j_m(q\xi)]'},\\
        b_m=&\frac{\mu_1j_m(q\xi)[\xi j_m(\xi)]'-j_m(\xi)[q\xi j_m(q\xi)]'}{\mu_1j_m(q\xi)[\xi h_m(\xi)]'-h_m(\xi)[q\xi j_m(q\xi)]'},
    \end{align}
\end{subequations}
where $q$ refers to the complex refractive index of the particle to the ambient medium and $\mu_1$ is the ratio of the magnetic permeability of the particle to the magnetic permeability of the ambient medium. 
For condensed particles, the scattering cross section  $\sigma_{S|M}$, called Mie scattering cross section, is given as~\cite{9783527618156},
\begin{subequations}
    \begin{align}
        \sigma_{S|M}=&\frac{\pi d^2}{2\xi^2}\sum_{m=1}^\infty\left(2m+1\right)\left(|a_m|^2+|b_m|^2\right)\\
        \approx&\frac{\pi d^2}{2\xi^2}\sum_{m=1}^{m_{\operatorname{max}}}\left(2m+1\right)\left(|a_m|^2+|b_m|^2\right).\label{eq:SM}
    \end{align}
\end{subequations}
The approximations in \eqref{eq:AM} and \eqref{eq:SM} are effective by satisfying the following condition~\cite{9783527618156},
\begin{equation}
    m_{\operatorname{max}}\ge\xi+4\xi^{1/3}+2.
\end{equation}
Mie absorption cross section and Mie scattering cross section depend on the chemical composition and diameter of particles, which are reflected in the complex refractive index $q$ and the variable $\xi$, respectively.

\begin{figure}[t]
    \centering
    \includegraphics[width=0.48\textwidth]{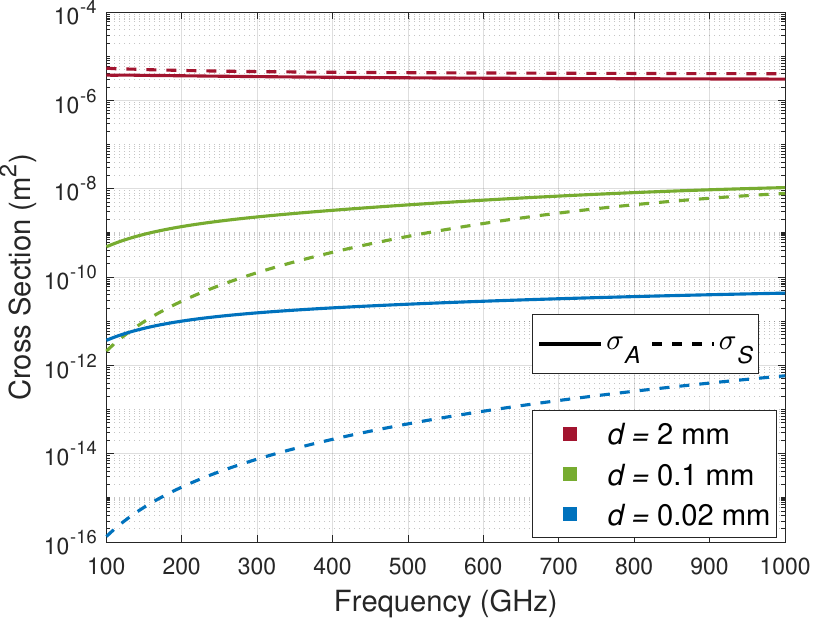}
		\captionsetup{font=footnotesize}
		\caption{The cross sections of water drops with various diameters.}
		\label{fig:MieCS}
\end{figure}

Even though \eqref{eq:AM} and \eqref{eq:SM} are applicable for particles with various diameters, they require high computational complexity.
Therefore, the geometric approximation or the Rayleigh approximation is employed to simplify the calculation of the cross sections with respect to the particles with diameter $d\gg\lambda$ or $d\ll\lambda$, while no further approximation is suitable for the particles with diameter $d\approx\lambda$.
Hence, we make a stipulation that Mie absorption and Mie scattering typically refer to the corresponding effect induced by the particles with diameter $d\approx\lambda$ in this paper.
Moreover, the geometric approximation will not be discussed in this paper since the diameter $d$ of the particles in the atmosphere are generally not much larger than the wavelength $\lambda$ of THz waves~\cite{10023042}.

Fig.~\ref{fig:MieCS} illustrates the absorption cross section and the scattering cross section of water drops with various diameters, where complex refractive index of water drops is acquired through the double-Debye model.
For THz waves from 0.1 to 1 THz, the collision between photons and water drops with $d=2$~mm, typical raindrops, induces Mie absorption and Mie scattering, of which the Mie absorption cross section and the Mie scattering cross section are both on the order of $10^{-6}$~m$^2$.
Moreover, the Mie scattering cross section is slightly larger than the Mie absorption cross section of the water drop with $d=2$~mm.
Therefore, THz waves are attenuated severer by the scattering effect rather than the absorption effect when the collision with raindrops occurs.

\subsection{Rayleigh Absorption and Scattering}

When the diameter of condensed particles is much smaller than $\lambda$, e.g. $d<\lambda/10$, Rayleigh approximation can be applied into \eqref{eq:AM} and \eqref{eq:SM}, in which $\xi$ to the lowest order are retained. 
Then, the Rayleigh absorption cross section, denoted by $\sigma_{A|R}$, is derived as~\cite{9783527618156},
\begin{equation}\label{eq:AR}
    \sigma_{A|R}=\frac{\pi d^3}{\lambda}\cdot\operatorname{Re}\left[i\left(\frac{q^2-1}{q^2+2}\right)\right],
\end{equation}
where $i$ is the imaginary unit, and the Rayleigh scattering cross section, denoted by $\sigma_{A|S}$, is~\cite{9783527618156},
\begin{equation}\label{eq:SR}
    \sigma_{S|R}=\frac{2\pi^5}{3}\cdot\frac{d^6}{\lambda^4}\cdot\left|\frac{q^2-1}{q^2+2}\right|^2.
\end{equation}
As illustrated in Fig.~\ref{fig:MieCS}, for water drops of $d=0.02$~mm, the Rayleigh scattering cross section is generally much smaller than the Rayleigh absorption cross section due to $d\ll\lambda$.
The cross sections of water drops of $d=0.1$~mm show the transition from Rayleigh theory to Mie theory as frequency increases and $\xi$ decreases.

\begin{figure}[t]
    \centering
    \includegraphics[width=0.48\textwidth]{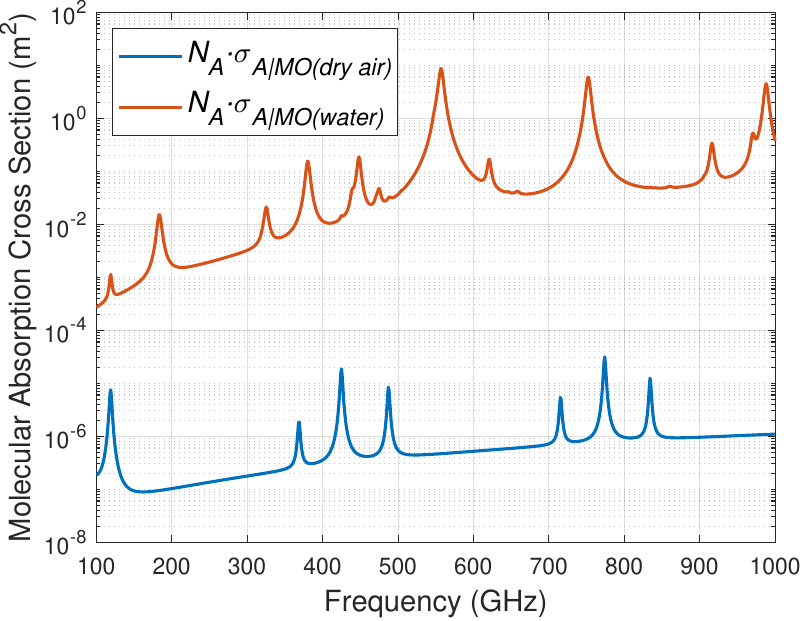}
		\captionsetup{font=footnotesize}
		\caption{The absorption cross section of molecules~in gaseous form, where $N_A$ is the Avogadro~constant.}
		\label{fig:MACS}
\end{figure}

\subsection{Molecular Absorption and Scattering}

When the particle is merely a single molecule, its cross sections cannot be acquired through \eqref{eq:AR} and \eqref{eq:SR}. 
On one hand, it is difficult to define the refractive index and diameter of a single molecule to absorb and scatter THz waves.
On the other hand, the absorption effect of molecules relies on the quantum property while the Mie and Rayleigh theories are based on classical electrodynamics.

In quantum mechanics, each kind of molecule in gaseous form has intrinsic and discrete rotational energy levels and is commonly at the lowest rotational energy level.
The difference between a high rotational energy level and the lowest one is defined as the rotational energy gap.
When the rotational energy gap of a molecule coincides with the energy of an incident photon, the photon will be absorbed and thus the EM wave is attenuated.
Meanwhile, the molecule is transferred from the lowest energy level to the high energy level.
The rotational energy gap of atmospheric molecules generally falls in the THz band and the millimeter wave, such as water vapor for the THz wave and oxygen for the millimeter wave~\cite{7731243}.
Therefore, molecular absorption is one of the most significant features of THz waves while it is negligible for the lower frequencies~\cite{9815184}. 
Fig.~\ref{fig:MACS} shows the molecular absorption cross section of water molecules and dry air molecules, where 1 mol dry air molecules consists of 0.78084 mol nitrogen, 0.20948 mol oxygen, 0.00935 mol argon and 0.00033 mol carbon dioxide~\cite{Bacal2008}.
The Rayleigh scattering cross section for a molecule, denoted by $\sigma_{S|MO}$, is expressed as,
\begin{equation}
    \sigma_{S|MO}=\frac{8\pi}{3}\cdot\left(\frac{2\pi}{\lambda}\right)^4\cdot\alpha^2,
\end{equation}
where $\alpha$ is the polarizability of the molecule. The polarizabilities of atmospheric molecules are all on the order of $10^{-30}$~m$^3$~\cite{Sharipov_2019,STAIKOVA2004213}.

It is worth noting that the molecular absorption cross section may vary with the state of matter even for the same type of molecules, since the interaction between different molecules becomes significant in the condensed state like the liquid state and changes the energy gap of the molecules.
Consequently, the intrinsic absorption peak disappear and the absorption progress is transitioned from quantum theory to classical theory.
Hence, the molecular absorption occurring in liquid or solid can be characterized by Rayleigh absorption or Mie absorption.

\begin{figure}[t]
    \centering
    \includegraphics[width=0.48\textwidth]{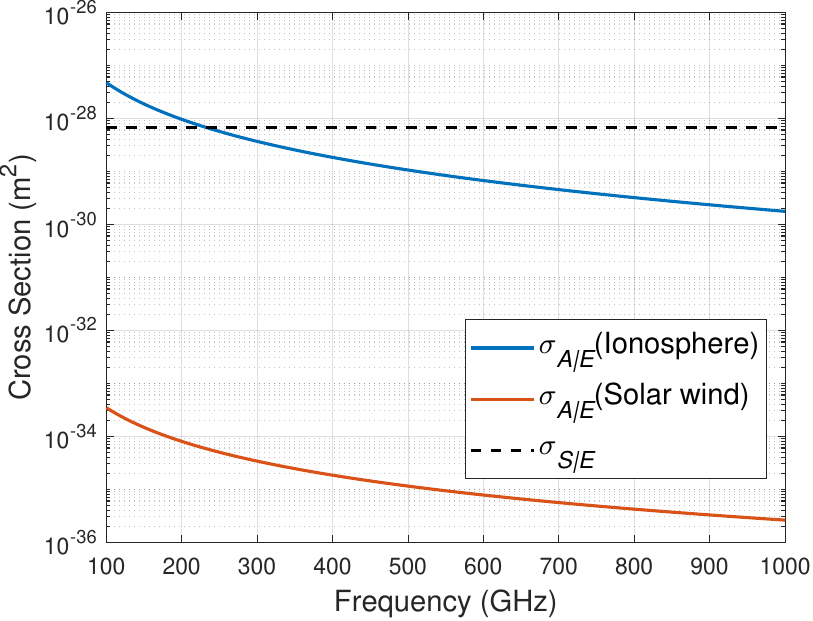}
		\captionsetup{font=footnotesize}
		\caption{The cross sections of free electrons in plasma.}
		\label{fig:ECS}
\end{figure}

\subsection{Coulomb Absorption and Thomson Scattering}

Furthermore, when the particles are free electrons in the ionized environment like plasma, the absorption effect, called Coulomb absorption, is characterized by Coulomb absorption cross section defined as~\cite{benzplasma},
\begin{equation}
    \sigma_{A|E}=\frac{\kappa_0}{n_e}\cdot\frac{f_{pe}^4}{f^2}\cdot\frac{1}{\sqrt{1-\frac{f_{pe}^2}{f^2}}},\label{eq:ACoul}
\end{equation}
where $f$ represents the frequency of the incident EM wave and $n_e$ is the number density of free electrons. The value $\kappa_0$ and the plasma frequency $f_{pe}$ are respectively given as~\cite{benzplasma},
\begin{equation}
    \kappa_0=1.76\times10^{-16}\left(\frac{\operatorname{ln}\Lambda_t}{T^{3/2}}\right)\left[\operatorname{m}^{-1}\operatorname{Hz}^{-2}\right],
\end{equation}

\begin{equation}
    f_{pe}=\frac{q_e}{2\pi}\sqrt{\frac{n_e}{\varepsilon_0m_e}},
\end{equation}
where $T$ in K refers to temperature of the environment, $\varepsilon_0$ represents the permittivity of free space, and $q_e$ and $m_e$ are the charge and the mass of a electron, respectively.
$\Lambda_t$ is the Gaunt factor given as~\cite{benzplasma},
\begin{equation}
    \resizebox{0.89\hsize}{!}{$\operatorname{ln}\Lambda_t=
    \left\{
    \begin{aligned}
        &38.3+\frac{3}{2}\operatorname{ln}T-\operatorname{ln}f&&\operatorname{for} T\lesssim 9.1\cdot 10^5 [\operatorname{K}].\\
        &45.3+\operatorname{ln}T-\operatorname{ln}f&&\operatorname{for} T\gtrsim 9.1\cdot 10^5 [\operatorname{K}].
    \end{aligned}
    \right.$}
\end{equation}
The scattering effect caused by free electrons is called Thomson scattering and characterized by the Thomson scattering cross section defined as~\cite{benzplasma},
\begin{equation}
    \sigma_{S|E}=\frac{8\pi}{3}\left(\frac{q_e^2}{4\pi\varepsilon_0m_ec^2}\right)^2,\label{eq:SThom}
\end{equation}
where $c$ is the speed of light in vacuum. 
The cross sections of free electrons are illustrated in Fig.~\ref{fig:ECS}.
The Coulomb absorption cross section in plasma decreases with the number density of free electrons and frequency, while the Thomson scattering cross section of free electrons shows no dependence as expected in~\eqref{eq:ACoul} and~\eqref{eq:SThom}.

\begin{figure}[t]
    \centering
    \includegraphics[width=0.48\textwidth,height=6.7cm]{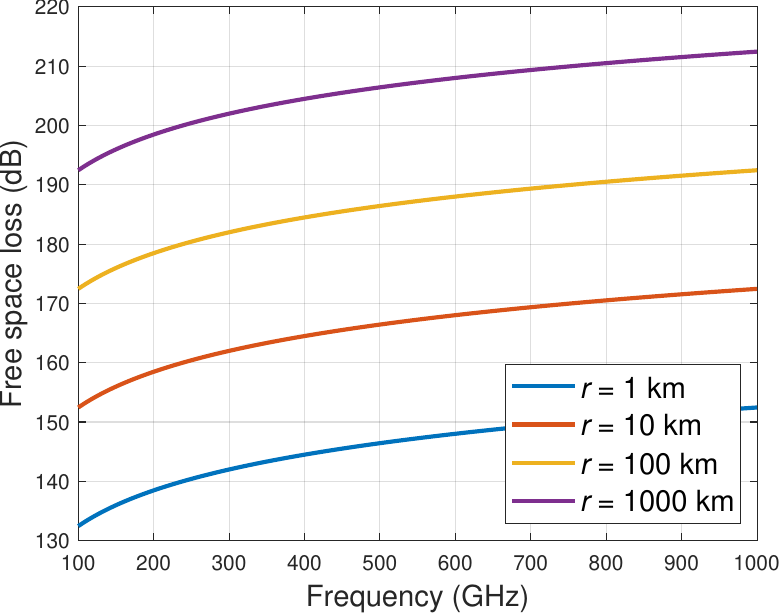}
		\captionsetup{font=footnotesize}
		\caption{The FSPL of THz waves with varying propagation distances.}
		\label{fig:FSPL}
\end{figure}

\subsection{Total Propagation Loss Model}
Even the propagation environment is a vacuum, due to the inverse-square law and the dependence of the effective area of antenna on frequency, EM wave still suffers from the FSPL defined~as,
\begin{equation}\label{eq:FSPL}
    L_{FP}=20\log_{10}\frac{4\pi r}{\lambda},
\end{equation}
where $r$ represents the path length and $L_{FP}$ is in the unit of dB.
Fig.~\ref{fig:FSPL} illustrates the FSPL of THz waves from 0.1 to 1~THz, which is higher than 150~dB for THz waves after propagation over 10~km.

Hence, by combining \eqref{eq:LC} and \eqref{eq:FSPL}, the total propagation loss is expressed as,
\begin{equation}\label{eq:Ltot}
    \begin{aligned}
        L_{tot}=&L_{C}+L_{FP}\\        =&10\log_{10}e\cdot\sum_{\varsigma}\left(\sigma_{A|\varsigma}+\sigma_{S|\varsigma}\right)\cdot\int_{s_0}^{s_1}n_{\varsigma}\left(s\right)\dif s\\&+20\log_{10}\frac{4\pi r}{\lambda},
    \end{aligned}
\end{equation}
where $L_{tot}$ is in the unit of dB and the cross sections $\sigma_{A|\varsigma},\sigma_{S|\varsigma}$ can be calculated from \eqref{eq:AM} to \eqref{eq:SThom}.
Based on the knowledge about the propagation path and the particles distribution $n_{\varsigma}$ along the path, the propagation loss of THz waves can be precisely assessed according to \eqref{eq:Ltot}.

\section{Numerical Evaluation of Propagation Loss}\label{sect:TPL}

From sea level to outer space, the atmosphere can be roughly divided into three layers, namely, the troposphere from 0 km to 10 km, the stratosphere from 10 km to 50 km and the ionosphere from 50 km to 1000~km.
As shown in Fig.~\ref{fig_Tpropa}, the particles in different layers are various and in nonuniform distribution.
Therefore, the propagation loss of THz waves in atmosphere cannot be assessed without knowledge on the spatial distribution of particles.
In this section, we first model the spatial distribution of common particles in atmosphere.
Then, by incorporating the spatial distribution model for the particles, we evaluate the propagation loss of a vertical THz link to illustrate each attenuation effect in different atmospheric layers.

\subsection{Particle Distribution}
The molecules except water molecules at the altitude $h$ is in gaseous form in the atmosphere without condensation.
On one hand, these molecules follow the hydrostatic equation expressed as~\cite{blundell2010concepts},
\begin{equation}\label{eq:hys}
    \dif p=-nm_0g\dif h,
\end{equation}
where $p$ is the pressure of the gas, $m_0$ denotes the mass of one single molecule and $g$ represents the gravitational acceleration.
On the other hand, since the gas in atmosphere follows the ideal gas law as~\cite{blundell2010concepts},
\begin{equation}\label{eq:igl}
    p=n(h)k_BT(h),
\end{equation}
where $k_B$ refers to the Boltzmann constant and the atmospheric temperature $T(h)$ can be modeled as,
\begin{equation}
    T(h)=\left\{
    \begin{aligned}
        &T(0)-\Gamma h,&&h\leq10^4\operatorname{m},\\
        &T(10^4),
        &&10^4\operatorname{m}<h\leq5\times10^4\operatorname{m},\\
        &2000\operatorname{K}, &&h>5\times10^4m,
    \end{aligned}
    \right.
\end{equation}
where the parameter $\Gamma$ is given by
\begin{equation}
    \Gamma=\left\{
    \begin{aligned}
        &0.006 \operatorname{K/m}, &h<10^4m,\\
        &0, &h>10^4m.
    \end{aligned}
    \right.
\end{equation}
Then, by differentiating the both sides of~\eqref{eq:igl}, $\dif p$ can be expressed as,
\begin{equation}\label{eq:difigl}
    \dif p=k_BT\dif n+nk_B\dif T=k_BT\dif n-nk_B\Gamma\dif h.
\end{equation}
By incorporating \eqref{eq:difigl} into \eqref{eq:hys}, we obtain a differential equation for $n$ as,
\begin{equation}\label{eq:difnda}
    \frac{\dif n}{n}=-\frac{m_0g-k_B\Gamma}{k_BT}\cdot\dif h,
\end{equation}
By integrating both sides of~\eqref{eq:difnda}, the number density $n$ versus altitude $h$ is given as,
\begin{equation}\label{eq:nda}
    n(h)=n(0)\exp\left\{-\frac{m_0g-k_B\Gamma}{k_B[T(0)-\Gamma h]}\cdot h\right\}.
\end{equation}
However, \eqref{eq:nda} cannot be applied to water vapor which could condense and only exist at the altitudes from 0 to 15~km due to the decrease of temperature~\cite{ITU835}.

For water vapor, since it may condense into water drops in humid atmosphere, the corresponding distribution should be characterized with respect to two conditions, saturated condition and unsaturated condition.
The condensation of water vapor can occur in the saturated condition while not in the unsaturated condition.
In the saturated condition, the relative humidity is assumed to be over 90\% at all altitudes such that the water vapor can condense.
By combining the ideal gas law and the Buck equation under the saturated condition~\cite{NewEquationsforComputingVP}, the number density of water molecules in gaseous form is derived as,
\begin{equation}\label{eq:nws}
    n_{SW|MO}(h)\approx\frac{550.09}{k_BT(h)}\cdot e^{\left[19.843-\frac{T(h)}{234.5}\right]\left[\frac{T(h)-273.15}{T(h)-16.01}\right]}.
\end{equation}
In the unsaturated condition, water vapor cannot condensed and the number density of water molecules can be modeled as~\cite{ITU835},
\begin{equation}\label{eq:nwu}
    n_{UW|MO}(h)=n_{UW|MO}(0)\cdot e^{-h/h_{W0}},
\end{equation}
where $h_{W0}=2000$ m.
By utilizing~\eqref{eq:nda}, \eqref{eq:nws}, and~\eqref{eq:nwu}, the spatial distribution of atmospheric molecules can be assessed.

For condensed particles in the atmosphere, such as raindrops, cloud droplets and fog droplets, their number densities are supposed to be independent of the altitude since they are in vertical motion or periodic evaporation-condensation.
Moreover, the condensed particles only exist in the troposphere as the condensation generally occurs in troposphere but not at higher altitudes~\cite{https://doi.org/10.1029/2009RG000301}.

\begin{figure}[t]
    \centering
    \includegraphics[width=0.48\textwidth]{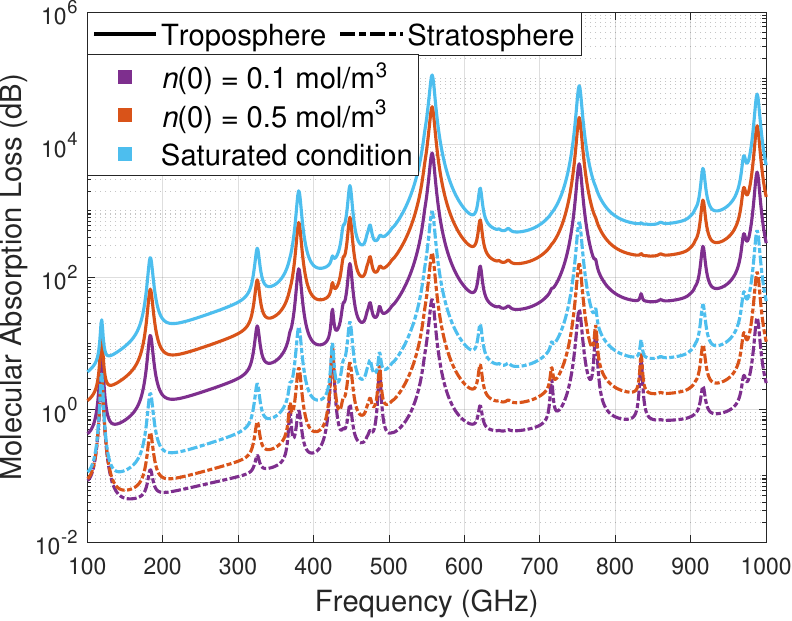}
	\captionsetup{font=footnotesize}
	\caption{The molecular absorption loss in the troposphere and stratosphere, where $T(0)=298.15$ K.}
	\label{fig:MATS}
\end{figure}
\begin{figure}[t]
    \centering
    \includegraphics[width=0.48\textwidth]{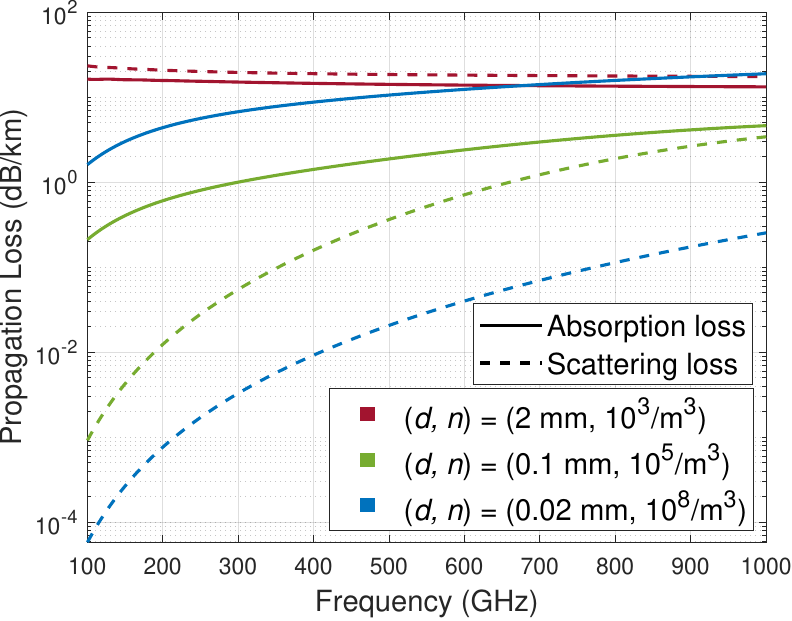}
    \captionsetup{font=footnotesize}
    \caption{The propagation loss caused by condensed particles in the troposphere.}
    \label{fig:CPLossT}
\end{figure}

\subsection{Propagation in the Troposphere}
With over half of the mass of the whole atmosphere and the abundant water vapor evaporated from the ocean~\cite{troposheremass}, the troposphere is the most harsh propagation environment for THz waves compared to the other layers of atmosphere.
Considering a vertical link across the whole troposphere, i.e., a link of 10 km with elevation angle $\theta=\pi/2$, the molecular absorption loss is shown in Fig.~\ref{fig:MATS}.
In the saturated condition, the THz waves above 320 GHz all suffer from severe molecular absorption loss over 100 dB.
In the unsaturated condition, such as $n(0)=0.5$ mol and $n(0)=0.1$ mol, the molecular absorption loss is much lower than that in the saturated condition due to fewer water vapor.
Moreover, the spectral windows, which are defined as the band with low molecular absorption loss, broaden with the decreasing number density of water molecules.
Unlike molecular absorption loss, the molecular scattering loss is on the order of $10^{-15}$ dB/km and thus negligible.
Since the density of atmospheric molecules decreases with the increasing altitude, the molecular scattering loss is also negligible in the upper atmosphere and will not be mentioned hereafter.

In addition to molecular absorption loss, THz waves may suffer from attenuation from condensed particles like water drops when water vapor condenses in the saturated condition.
The absorption and scattering loss caused by condensed particles is illustrated in Fig.~\ref{fig:CPLossT}.
In the heavy rain with a rain rate of $98$~mm/hr, the diameter and number density of typical raindrops are 2 mm and 10$^3$ m$^{-3}$, respectively~\cite{WALLACE2006271}.
For raindrops, the Mie scattering loss is greater than the Mie absorption loss and both of them decrease with an increasing frequency.
In heavy fog, the diameter and number density of typical fog droplets are 0.02 mm and 10$^8$ m$^{-3}$, respectively~\cite{WALLACE2006271}.
As are cloud droplets in cumulonimbus, a towering vertical cloud associated with heavy rain~\cite{WALLACE2006271}.
Different from the rain loss, the absorption loss of fog and cloud is much larger than the scattering loss and both of them increase with frequency.

\subsection{Propagation in the Stratosphere}
Due to the gravity and the decrease of temperature, there is much fewer water vapor in the stratosphere and thus much smaller molecular absorption loss than in the troposphere as demonstrated in Fig.~\ref{fig:MATS}.
Therefore, the spectral windows of THz waves in the stratosphere are broader than that in the troposphere.
Moreover, there are no condensed particles or free electrons so that there is no additional collisional loss in the stratosphere.
However, since water vapor still exists at the altitudes from 10 km to 15 km, the frequency close to the absorption peak should not be adopted for THz links crossing these region to avoid high molecular absorption loss.

\begin{figure}[t]
    \centering
    \includegraphics[width=0.48\textwidth]{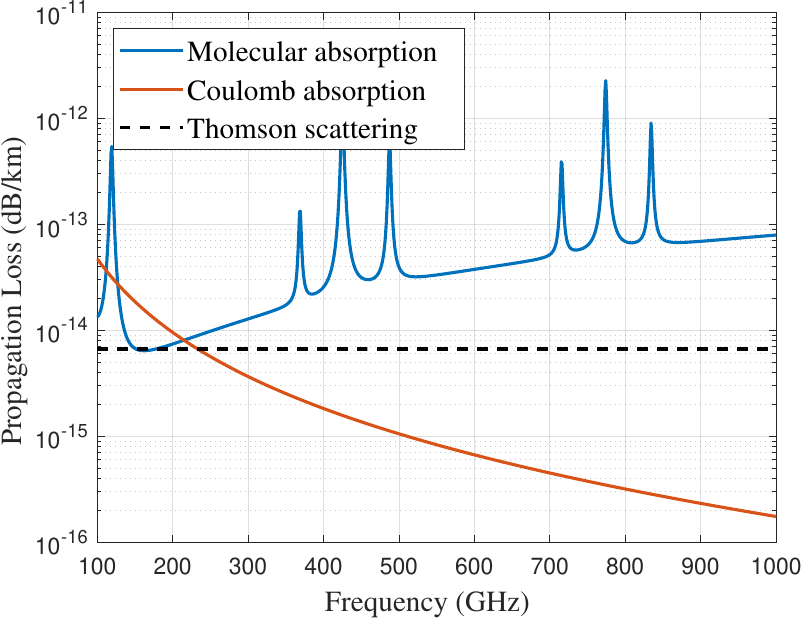}
	\captionsetup{font=footnotesize}
	\caption{The collisional loss in the ionosphere.}
	\label{fig:PLI}
\end{figure}

\subsection{Propagation in the Ionosphere}
In the ionosphere, there are free electrons ionized from neural molecules since the high-energy ray from the Sun and space is absorbed. 
The highest density of free electrons in the ionosphere is on the order of $10^{11}$ m$^{-3}$ and the typical temperature of ionosphere is 2000~K~\cite{9541155}. 
Therefore, the Coulomb absorption loss and Thomson scattering loss for THz waves in the ionosphere are both smaller than 10$^{-13}$~dB/km.
Fig.~\ref{fig:PLI} illustrates the collisional loss including the loss caused by molecular absorption, Coulomb absorption, and Thomson scattering for THz waves in the ionosphere.
The molecular absorption loss is on the order of $10^{-13}$ dB/km due to the sparse atmospheric molecules in the ionosphere.
Hence, the collisional loss of THz waves in the ionosphere is less than 10$^{-11}$~dB/km.
Compared with the FSPL over 150~dB, the collisional loss of THz waves in the ionosphere is negligible.
Therefore, the FSPL dominates the propagation loss of THz waves in the ionosphere.

\subsection{Propagation in the Outer Space}

Free electrons in solar wind are the most common particles that collides with THz waves in the outer space. 
Their typical temperature and density are $10^5$ K and 10$^7$ m$^{-3}$, respectively~\cite{9541155}. 
According to~\eqref{eq:ACoul} and~\eqref{eq:SThom}, the collisional loss of THz waves is less than 10$^{-24}$~dB/km.
Hence, the dominant attenuation effect for THz waves is FSPL in the outer space as in the ionosphere.

Based on the knowledge of the particles distribution and the corresponding cross sections, the propagation loss of THz waves in the atmosphere can be numerically evaluated.
Fig.~\ref{fig_Tpropa} summarizes the evaluation of propagation loss for THz waves in the atmosphere.
In the troposphere, the attenuation effects including FSPL, molecular absorption, Rayleigh absorption, Mie absorption and scattering can lead to significant loss on THz waves.
Specifically, the loss caused by Rayleigh absorption, Mie absorption and scattering are all on the order of 10$^{1}$~dB/km, while that of Rayleigh scattering is 10$^{-1}$~dB/km.
Molecular absorption loss is extremely high at the frequencies close to absorption peaks and relatively low in the spectral window.
Therefore, the frequencies adopted in THz links in the troposphere should fall in the spectral window to avoid severe molecular absorption loss.
In the stratosphere, FSPL is generally the most significant attenuation effect except for the frequencies close to absorption peaks.
At the altitudes higher than 50~km, namely, the ionosphere and the outer space, the attenuation effects except FSPL are all negligible as they are all less than 10$^{-11}$~dB/km.

\section{Improvement on Existing ITU Models}\label{sec:comitu}
The ITU recommendation models are commonly utilized to predict the propagation loss of the THz waves from 0.1 to 1 THz in the troposphere~\cite{ITU676,ITU838,ITU840}.
The ITU models are easy-to-use and they balance the computational complexity and precision.
However, the ITU models for THz waves are empirical, which is a double-edged sword.
On one hand, the empirical model is less complex compared with theoretical model.
On the other hand, since empirical models are developed from the data measured in a typical environment, they neglect the difference of various environment and result in prediction error.
Moreover, the ITU models characterize only a certain portions of attenuation effects while the model proposed in this paper can incorporate the common attenuation effects for THz waves in the atmosphere.
In this section, we make a comparison between the proposed model with ITU-R P.838-3 and ITU-R P.676-13 with respect to the propagation loss caused by rain and molecular absorption loss.
However, there is no comparison between the proposed model and ITU-R P.840-8 since both of them are based on Rayleigh theory and thus they fit well with each other.

\begin{figure}[t]
    \centering
    \includegraphics[width=0.48\textwidth]{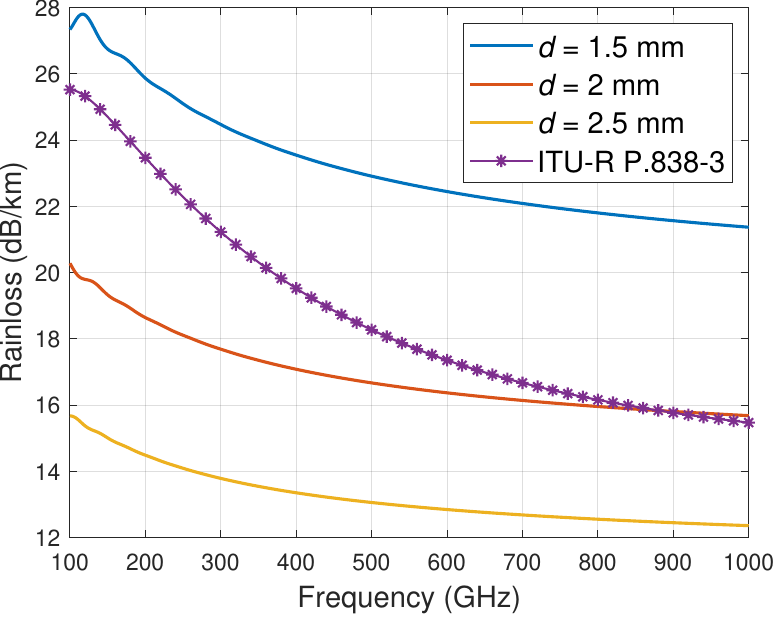}
    \captionsetup{font=footnotesize}
    \caption{The prediction of rain attenuation by the proposed model and ITU-R P.838-3, where the rain is supposed to only consist of raindrops with diameter~$d$ and the rain rate is 50 mm/hr.}
    \label{fig:RLC}
\end{figure}

\subsection{Improvement of ITU-R P.838-3 on Rain Attenuation}
In general, the diameter of raindrops is not a invariant constant and the size distribution of raindrops is called raindrop spectrum.
There are various raindrop spectra even for the rain of the same rain rate.
For instance, the rain above sea consists of raindrop with smaller diameter and higher number density compared with the rain above land~\cite{10023042}.
Consequently, the propagation loss caused by rain above sea is different from that by rain above land, while ITU-R P.838-3 only incorporates the impact of rain rate and thus cannot distinguish the different rain spectra.

As illustrated in Fig.~\ref{fig:RLC}, the rain consisting of raindrops with smaller diameter can induce severer attenuation than that of raindrops with larger diameter of the same rain rate.
From~\eqref{eq:LC}, ~\eqref{eq:AM}, and~\eqref{eq:SM},
we can conclude the following relation, namely, $n\propto d^{-3}$, $\sigma_{A|M},\sigma_{S|M}\propto d^{2}$, $L_M\propto n\cdot(\sigma_{A|M}+\sigma_{S|M})\propto d^{-1}$.
For rain with a given rain rate, the absorption and scattering loss increase as the average diameter of raindrops decreases.
Therefore, THz waves suffer from higher propagation loss in the rain above sea than the rain above land, which cannot be demonstrated by ITU-R P.838-3.
Meanwhile, the proposed model can incorporate various rain spectra through varying the input of $n_\xi$ in~\eqref{eq:Ltot} and thus capture the difference of various environment.

\textit{Next step for standardization}: Hence, in the case without sufficient data on rain spectra, ITU-R P.838-3 can provide a preliminary estimation on the attenuation.
In the case with knowledge on rain spectra, the proposed model can more accurately characterize the propagation loss of THz waves through the computation from~\eqref{eq:AM} to~\eqref{eq:SR} and~\eqref{eq:Ltot}.

\begin{figure}[t]
    \centering
    \includegraphics[width=0.48\textwidth]{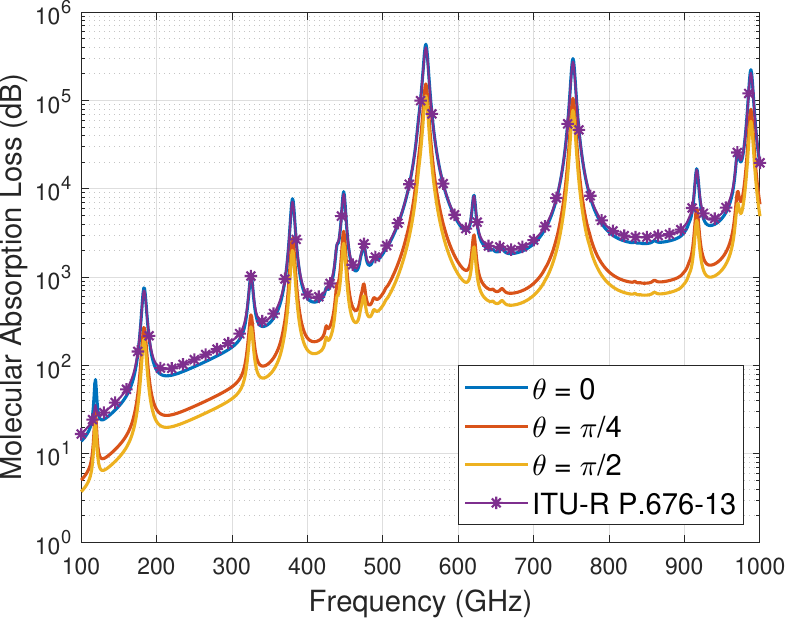}
    \captionsetup{font=footnotesize}
    \caption{The molecular absorption loss with various elevation angles at $0$, $\pi/4$, and $\pi/2$, based on the proposed model under saturated condition, where the propagation distance is 10 km and T(0)=298.15 K.}
    \label{fig:MALC}
\end{figure}





\subsection{Improvement of ITU-R P.676-13 on Molecular Absorption Loss}
As discussed in Sec.~\ref{sect:TPL}, the spatial distribution of atmospheric molecules is nonuniform and the number density decreases with the increasing altitude.
Ignoring inhomogeneity of atmospheric molecules could lead to severe prediction error of THz propagation loss. 
As illustrated in Fig.~\ref{fig:MALC}, the difference of molecular absorption loss can reach several hundreds dB for links with various elevation angle $\theta$ though the propagation distance is 10~km for all cases.
The prediction of molecular absorption loss given by the proposed model is the same as ITU-R P.676-13 at the absorption peaks and slightly lower in the regions between peaks.
This is caused by the minor difference in the characterization on the amount of molecules.
ITU-R P.676-13 characterizes the amount of molecules through pressure $p$ while the proposed model use number density $n$.
There are more than one formulas, such as ideal gas law and Van der Waals equation~\cite{KLEIN197428}, to compute the conversion between $p$ and $n$ and the conversion results of various formulas are slightly different.
Both the proposed model and ITU-R P.676-13 can characterize the effect of the inhomogeneity of molecules which is featured by number density in the former model and partial pressure in the latter one.
However, ITU-R P.676-13 is only useful to evaluate the molecular absorption.
By contrast, the proposed model has higher interpretability and can incorporate more attenuation effects including Mie absorption and scattering, Rayleigh absorption and scattering, Coulomb absorption and Thomson scattering.

\textit{Next step for standardization:} Hence, ITU-R P.676-13 can handle THz wave propagation in the environment where no condensed particles exist, while it cannot cope with that in foggy weather or ionized environment. A thorough incorporation of these weather effects are needed, by invoking the equations from~\eqref{eq:AM} to~\eqref{eq:Ltot} of this work.


\begin{table*}[ht]
\centering
\caption{Frequency allocations from 0.1 to 0.45 THz~\cite{WRC95to19}}
\label{tab:band}
\begin{threeparttable}
\begin{tabular}{|c|c|c|cc|cc|}
\hline
 &  &  & \multicolumn{2}{c|}{} & \multicolumn{2}{c|}{} \\
 &  &  & \multicolumn{2}{c|}{\multirow{-2}{*}{Fiexed-satellite service}} & \multicolumn{2}{c|}{\multirow{-2}{*}{Mobile-satellite service}} \\ \cline{4-7} 
 &  &  & \multicolumn{1}{c|}{} &  & \multicolumn{1}{c|}{} &  \\
\multirow{-4}{*}{Frequencies (GHz)} & \multirow{-4}{*}{Fixed and mobile service} & \multirow{-4}{*}{Aeronautical mobile service} & \multicolumn{1}{c|}{\multirow{-2}{*}{Earth to space}} & \multirow{-2}{*}{Space to earth} & \multicolumn{1}{c|}{\multirow{-2}{*}{Earth to space}} & \multirow{-2}{*}{Space to earth} \\ \hline
 &  &  & \multicolumn{1}{c|}{} &  & \multicolumn{1}{c|}{} &  \\
\multirow{-2}{*}{111.8-114.25} & \multirow{-2}{*}{\checkmark} & \multirow{-2}{*}{} & \multicolumn{1}{c|}{\multirow{-2}{*}{}} & \multirow{-2}{*}{} & \multicolumn{1}{c|}{\multirow{-2}{*}{}} & \multirow{-2}{*}{} \\ \hline
 &  &  & \multicolumn{1}{c|}{} &  & \multicolumn{1}{c|}{} &  \\
\multirow{-2}{*}{122.25-123} & \multirow{-2}{*}{\checkmark} & \multirow{-2}{*}{\checkmark} & \multicolumn{1}{c|}{\multirow{-2}{*}{}} & \multirow{-2}{*}{} & \multicolumn{1}{c|}{\multirow{-2}{*}{}} & \multirow{-2}{*}{} \\ \hline
{\color[HTML]{FE0000} } &  &  & \multicolumn{1}{c|}{} & {\color[HTML]{FE0000} } & \multicolumn{1}{c|}{} & {\color[HTML]{FE0000} } \\
\multirow{-2}{*}{{\color[HTML]{FE0000} \textbf{123-130}}} & \multirow{-2}{*}{} & \multirow{-2}{*}{} & \multicolumn{1}{c|}{\multirow{-2}{*}{}} & \multirow{-2}{*}{{\color[HTML]{FE0000} \textbf{\ding{52}}}} & \multicolumn{1}{c|}{\multirow{-2}{*}{}} & \multirow{-2}{*}{{\color[HTML]{FE0000} \textbf{\ding{52}}}} \\ \hline
{\color[HTML]{FE0000} } &  & {\color[HTML]{FE0000} } & \multicolumn{1}{c|}{} &  & \multicolumn{1}{c|}{} &  \\
\multirow{-2}{*}{{\color[HTML]{FE0000} \textbf{130-134}}} & \multirow{-2}{*}{\checkmark} & \multirow{-2}{*}{{\color[HTML]{FE0000} \textbf{\ding{52}}}} & \multicolumn{1}{c|}{\multirow{-2}{*}{}} & \multirow{-2}{*}{} & \multicolumn{1}{c|}{\multirow{-2}{*}{}} & \multirow{-2}{*}{} \\ \hline
{\color[HTML]{FE0000} } & {\color[HTML]{FE0000} } &  & \multicolumn{1}{c|}{} &  & \multicolumn{1}{c|}{} &  \\
\multirow{-2}{*}{{\color[HTML]{FE0000} \textbf{141-148.5}}} & \multirow{-2}{*}{{\color[HTML]{FE0000} \textbf{\ding{52}}}} & \multirow{-2}{*}{} & \multicolumn{1}{c|}{\multirow{-2}{*}{}} & \multirow{-2}{*}{} & \multicolumn{1}{c|}{\multirow{-2}{*}{}} & \multirow{-2}{*}{} \\ \hline
{\color[HTML]{FE0000} } & {\color[HTML]{FE0000} } &  & \multicolumn{1}{c|}{} &  & \multicolumn{1}{c|}{} &  \\
\multirow{-2}{*}{{\color[HTML]{FE0000} \textbf{151.5-158.5}}} & \multirow{-2}{*}{{\color[HTML]{FE0000} \textbf{\ding{52}}}} & \multirow{-2}{*}{} & \multicolumn{1}{c|}{\multirow{-2}{*}{}} & \multirow{-2}{*}{} & \multicolumn{1}{c|}{\multirow{-2}{*}{}} & \multirow{-2}{*}{} \\ \hline
{\color[HTML]{FE0000} } & {\color[HTML]{FE0000} } &  & \multicolumn{1}{c|}{} &  & \multicolumn{1}{c|}{} &  \\
\multirow{-2}{*}{{\color[HTML]{FE0000} \textbf{158.5-164}}} & \multirow{-2}{*}{{\color[HTML]{FE0000} \textbf{\ding{52}}}} & \multirow{-2}{*}{} & \multicolumn{1}{c|}{\multirow{-2}{*}{}} & \multirow{-2}{*}{\checkmark} & \multicolumn{1}{c|}{\multirow{-2}{*}{}} & \multirow{-2}{*}{\checkmark} \\ \hline
{\color[HTML]{FE0000} } &  & {\color[HTML]{FE0000} } & \multicolumn{1}{c|}{} &  & \multicolumn{1}{c|}{} &  \\
\multirow{-2}{*}{{\color[HTML]{FE0000} \textbf{167-174.5}}} & \multirow{-2}{*}{\checkmark} & \multirow{-2}{*}{{\color[HTML]{FE0000} \textbf{\ding{52}}}} & \multicolumn{1}{c|}{\multirow{-2}{*}{}} & \multirow{-2}{*}{\checkmark} & \multicolumn{1}{c|}{\multirow{-2}{*}{}} & \multirow{-2}{*}{} \\ \hline
{\color[HTML]{FE0000} } &  & {\color[HTML]{FE0000} } & \multicolumn{1}{c|}{} &  & \multicolumn{1}{c|}{} &  \\
\multirow{-2}{*}{{\color[HTML]{FE0000} \textbf{174.5-174.8}}} & \multirow{-2}{*}{\checkmark} & \multirow{-2}{*}{{\color[HTML]{FE0000} \textbf{\ding{52}}}} & \multicolumn{1}{c|}{\multirow{-2}{*}{}} & \multirow{-2}{*}{} & \multicolumn{1}{c|}{\multirow{-2}{*}{}} & \multirow{-2}{*}{} \\ \hline
 &  &  & \multicolumn{1}{c|}{} &  & \multicolumn{1}{c|}{} &  \\
\multirow{-2}{*}{191.8-200} & \multirow{-2}{*}{\checkmark} & \multirow{-2}{*}{\checkmark} & \multicolumn{1}{c|}{\multirow{-2}{*}{}} & \multirow{-2}{*}{} & \multicolumn{1}{c|}{\multirow{-2}{*}{\checkmark}} & \multirow{-2}{*}{\checkmark} \\ \hline
{\color[HTML]{FE0000} } & {\color[HTML]{FE0000} } &  & \multicolumn{1}{c|}{{\color[HTML]{FE0000} }} &  & \multicolumn{1}{c|}{} &  \\
\multirow{-2}{*}{{\color[HTML]{FE0000} \textbf{209-226}}} & \multirow{-2}{*}{{\color[HTML]{FE0000} \textbf{\ding{52}}}} & \multirow{-2}{*}{} & \multicolumn{1}{c|}{\multirow{-2}{*}{{\color[HTML]{FE0000} \textbf{\ding{52}}}}} & \multirow{-2}{*}{} & \multicolumn{1}{c|}{\multirow{-2}{*}{}} & \multirow{-2}{*}{} \\ \hline
 &  &  & \multicolumn{1}{c|}{} &  & \multicolumn{1}{c|}{} &  \\
\multirow{-2}{*}{231.5-232} & \multirow{-2}{*}{\checkmark} & \multirow{-2}{*}{} & \multicolumn{1}{c|}{\multirow{-2}{*}{}} & \multirow{-2}{*}{} & \multicolumn{1}{c|}{\multirow{-2}{*}{}} & \multirow{-2}{*}{} \\ \hline
 &  &  & \multicolumn{1}{c|}{} &  & \multicolumn{1}{c|}{} &  \\
\multirow{-2}{*}{232-235} & \multirow{-2}{*}{\checkmark} & \multirow{-2}{*}{} & \multicolumn{1}{c|}{\multirow{-2}{*}{}} & \multirow{-2}{*}{\checkmark} & \multicolumn{1}{c|}{\multirow{-2}{*}{}} & \multirow{-2}{*}{} \\ \hline
 &  &  & \multicolumn{1}{c|}{} &  & \multicolumn{1}{c|}{} &  \\
\multirow{-2}{*}{235-238} & \multirow{-2}{*}{} & \multirow{-2}{*}{} & \multicolumn{1}{c|}{\multirow{-2}{*}{}} & \multirow{-2}{*}{\checkmark} & \multicolumn{1}{c|}{\multirow{-2}{*}{}} & \multirow{-2}{*}{} \\ \hline
 &  &  & \multicolumn{1}{c|}{} &  & \multicolumn{1}{c|}{} &  \\
\multirow{-2}{*}{238-240} & \multirow{-2}{*}{\checkmark} & \multirow{-2}{*}{} & \multicolumn{1}{c|}{\multirow{-2}{*}{}} & \multirow{-2}{*}{\checkmark} & \multicolumn{1}{c|}{\multirow{-2}{*}{}} & \multirow{-2}{*}{} \\ \hline
 &  &  & \multicolumn{1}{c|}{} &  & \multicolumn{1}{c|}{} &  \\
\multirow{-2}{*}{240-241} & \multirow{-2}{*}{\checkmark} & \multirow{-2}{*}{} & \multicolumn{1}{c|}{\multirow{-2}{*}{}} & \multirow{-2}{*}{} & \multicolumn{1}{c|}{\multirow{-2}{*}{}} & \multirow{-2}{*}{} \\ \hline
{\color[HTML]{FE0000} } &  &  & \multicolumn{1}{c|}{} &  & \multicolumn{1}{c|}{{\color[HTML]{FE0000} }} &  \\
\multirow{-2}{*}{{\color[HTML]{FE0000} \textbf{252-265}}} & \multirow{-2}{*}{\checkmark} & \multirow{-2}{*}{} & \multicolumn{1}{c|}{\multirow{-2}{*}{}} & \multirow{-2}{*}{} & \multicolumn{1}{c|}{\multirow{-2}{*}{{\color[HTML]{FE0000} \textbf{\ding{52}}}}} & \multirow{-2}{*}{} \\ \hline
 &  &  & \multicolumn{1}{c|}{} &  & \multicolumn{1}{c|}{} &  \\
\multirow{-2}{*}{265-275} & \multirow{-2}{*}{\checkmark} & \multirow{-2}{*}{} & \multicolumn{1}{c|}{\multirow{-2}{*}{\checkmark}} & \multirow{-2}{*}{} & \multicolumn{1}{c|}{\multirow{-2}{*}{}} & \multirow{-2}{*}{} \\ \hline
 &  &  & \multicolumn{1}{c|}{} &  & \multicolumn{1}{c|}{} &  \\
\multirow{-2}{*}{275-296} & \multirow{-2}{*}{\checkmark} & \multirow{-2}{*}{} & \multicolumn{1}{c|}{\multirow{-2}{*}{}} & \multirow{-2}{*}{} & \multicolumn{1}{c|}{\multirow{-2}{*}{}} & \multirow{-2}{*}{} \\ \hline
 &  &  & \multicolumn{1}{c|}{} &  & \multicolumn{1}{c|}{} &  \\
\multirow{-2}{*}{306-313} & \multirow{-2}{*}{\checkmark} & \multirow{-2}{*}{} & \multicolumn{1}{c|}{\multirow{-2}{*}{}} & \multirow{-2}{*}{} & \multicolumn{1}{c|}{\multirow{-2}{*}{}} & \multirow{-2}{*}{} \\ \hline
 &  &  & \multicolumn{1}{c|}{} &  & \multicolumn{1}{c|}{} &  \\
\multirow{-2}{*}{318-333} & \multirow{-2}{*}{\checkmark} & \multirow{-2}{*}{} & \multicolumn{1}{c|}{\multirow{-2}{*}{}} & \multirow{-2}{*}{} & \multicolumn{1}{c|}{\multirow{-2}{*}{}} & \multirow{-2}{*}{} \\ \hline
 &  &  & \multicolumn{1}{c|}{} &  & \multicolumn{1}{c|}{} &  \\
\multirow{-2}{*}{356-450} & \multirow{-2}{*}{\checkmark} & \multirow{-2}{*}{} & \multicolumn{1}{c|}{\multirow{-2}{*}{}} & \multirow{-2}{*}{} & \multicolumn{1}{c|}{\multirow{-2}{*}{}} & \multirow{-2}{*}{} \\ \hline
\end{tabular}
\begin{tablenotes}
    \footnotesize  
    \item[] The frequencies not listed in this table are not allocated for fixed and mobile service, aeronautical mobile service, fixed-satellite service, or mobile-satellite service.
\end{tablenotes}
\end{threeparttable}
\end{table*}

\begin{table*}[ht]
\centering
\caption{THz space-ground uplink}
\label{tab:spgup}
\begin{threeparttable}
\resizebox{\textwidth}{!}{%
\begin{tabular}{cccccccccccc}
Link & $f$ (GHz) & $P_T$ (W) & $L_{FP}$ (dB) & $L_{M}$ (dB) & $L_R$ (dB) & $L_{MO}$ (dB) & $L_{tot}$ (dB) & $P_R$ (dBm) & $P_N$ (dBm) & $C$ (Gbps) & $C/B$ (bps/Hz) \\ \hline
NRS & 209-226 & 10 & 194.00 & 0 & 0 & 20.50 & 214.50 & -64.50 & -71.70 & 34.35 & 2.64 \\ \hline
NRR & 209-226 & 10 & 194.00 & 18.45 & 24.17 & 20.50 & 257.12 & -107.12 & -71.70 & 0.01 & 0.00 \\ \hline
La1 & 130-134 & 9.90 & 154.86 & 19.70 & 12.53 & 6.67 & 193.76 & -43.80 & -77.98 & \multirow{2}{*}{45.41} & 11.35 \\
Lb1 & 209-226 & 0.10 & 193.84 & 0 & 0 & 0.23 & 194.07 & -64.07 & -71.70 &  & 3.61 \\ \hline
La1 & 167-174.8 & 9.96 & 157.10 & 19.05 & 18.03 & 20.37 & 214.55 & -64.57 & -75.08 & \multirow{2}{*}{28.20} & 3.61 \\
Lb1 & 209-226 & 0.04 & 193.84 & 0 & 0 & 0.23 & 194.07 & -68.05 & -71.70 &  & 2.26 \\ \hline
La2 & 130-134 & 9.91 & 157.85 & 27.86 & 0 & 9.43 & 195.13 & -45.17 & -77.98 & \multirow{2}{*}{43.60} & 10.90 \\
Lb2 & 209-226 & 0.09 & 193.84 & 0 & 0 & 0.23 & 194.07 & -64.53 & -71.70 &  & 3.45 \\ \hline
La2 & 167-174.8 & 9.96 & 160.09 & 26.94 & 0 & 28.81 & 215.83 & -65.85 & -75.08 & \multirow{2}{*}{25.18} & 3.23 \\
Lb2 & 209-226 & 0.04 & 193.84 & 0 & 0 & 0.23 & 194.07 & -68.05 & -71.70 &  & 2.26
\end{tabular}%
}
\begin{tablenotes}
    \footnotesize  
    \item[] NRS and NRR denote the link without relay nodes in sunny and rainy weather, respectively.
\end{tablenotes}
\end{threeparttable}
\end{table*}

\begin{table*}[ht]
\centering
\caption{THz space-ground downlink}
\label{tab:spgdown}
\resizebox{\textwidth}{!}{%
\begin{tabular}{cccccccccccc}
Link & $f$ (GHz) & $P_T$ (W) & $L_{FP}$ (dB) & $L_{M}$ (dB) & $L_R$ (dB) & $L_{MO}$ (dB) & $L_{tot}$ (dB) & $P_R$ (dBm) & $P_N$ (dBm) & $C$ (Gbps) & $C/B$ (bps/Hz) \\ \hline
NRS & 123-130 & 10 & 189.30 & 0 & 0 & 7.11 & 196.41 & -46.41 & -75.55 & 67.77 & 9.68 \\ \hline
NRR & 123-130 & 10 & 189.30 & 19.76 & 11.74 & 7.11 & 227.91 & -77.91 & -75.55 & 4.62 & 0.66 \\ \hline
La1 & 130-134 & 9.78 & 154.86 & 19.70 & 12.53 & 6.67 & 193.76 & -43.86 & -77.98 & \multirow{2}{*}{45.34} & 11.34 \\
Lb1 & 123-130 & 0.22 & 189.14 & 0 & 0 & 0.30 & 189.44 & -56.02 & -75.55 &  & 6.50 \\ \hline
La1 & 167-174.8 & 9.96 & 157.10 & 19.05 & 18.03 & 20.37 & 214.55 & -64.57 & -75.08 & \multirow{2}{*}{28.20} & 3.61 \\
Lb1 & 123-130 & 0.04 & 189.14 & 0 & 0 & 0.30 & 189.44 & -63.42 & -75.55 &  & 4.12 \\ \hline
La2 & 130-134 & 9.81 & 157.85 & 27.86 & 0 & 9.43 & 195.13 & -45.22 & -77.98 & \multirow{2}{*}{43.54} & 10.88 \\
Lb2 & 123-130 & 0.19 & 189.14 & 0 & 0 & 0.30 & 189.44 & -56.65 & -75.55 &  & 6.30 \\ \hline
La2 & 167-174.8 & 9.97 & 160.09 & 26.94 & 0 & 28.81 & 215.83 & -65.85 & -75.08 & \multirow{2}{*}{25.19} & 3.23 \\
Lb2 & 123-130 & 0.03 & 189.14 & 0 & 0 & 0.30 & 189.44 & -64.67 & -75.55 &  & 3.73
\end{tabular}%
}
\end{table*}

\section{Evaluation of THz SAGIN Capacity}\label{sec:CEstim}
\begin{figure}[t]
    \centering
    \includegraphics[width=0.48\textwidth]{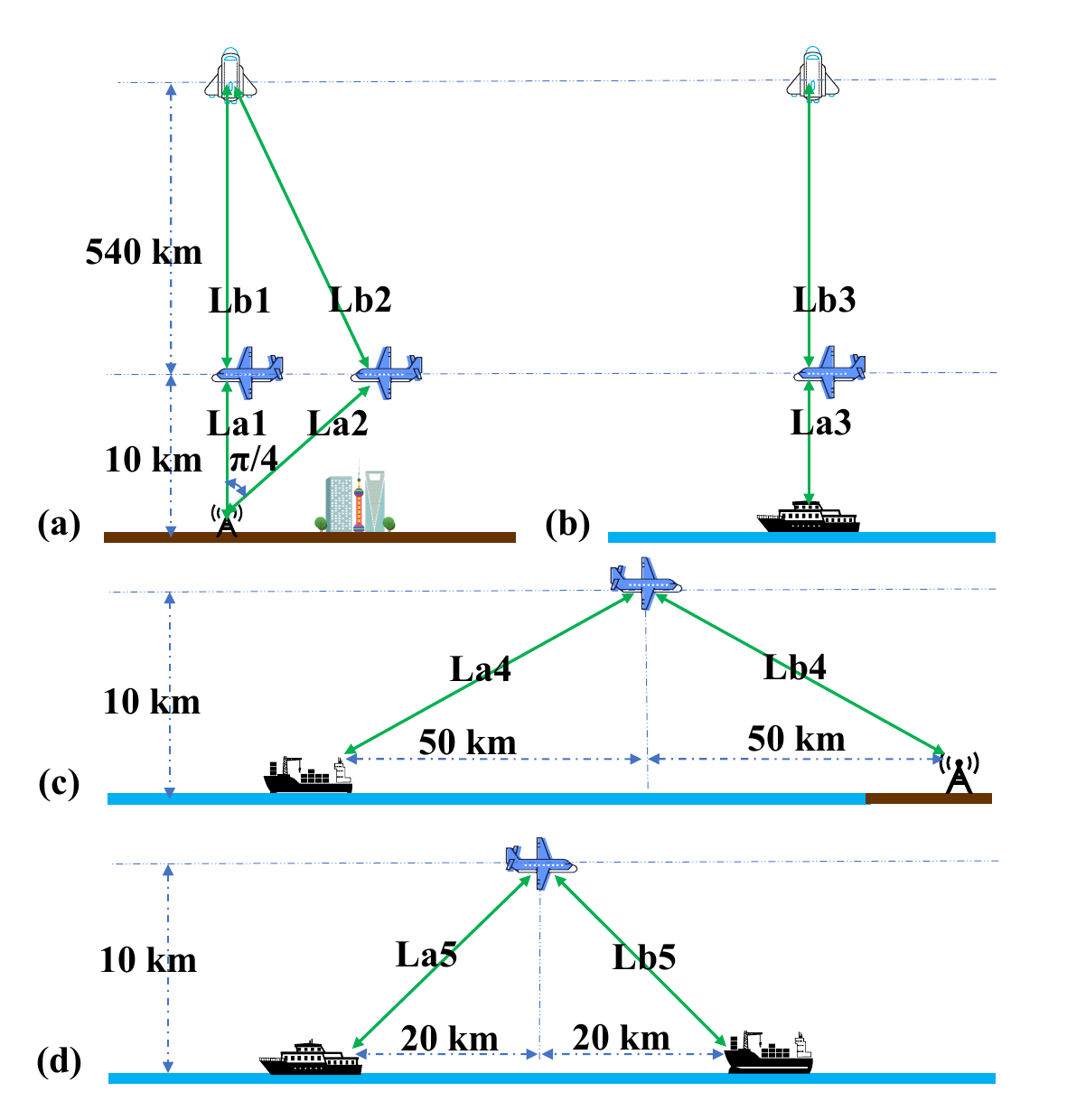}
    \captionsetup{font=footnotesize}
    \caption{A THz SAGIN including (a) space-ground, (b) space-sea, (c) ground-air-sea, and (d) sea-air-sea links.}
    \label{fig_relay}
\end{figure}

Based on the knowledge on the propagation loss of THz waves in the atmosphere, the performance of THz SAGIN can be estimated by performing the link budget analysis.
Specifically, ITU Radio Regulations restrict the applicable services for various frequency bands.
TABLE~\ref{tab:band} lists the frequency allocations from 0.1 to 0.45~THz, i.e., the option of carrier frequencies in THz SAGIN~\cite{WRC95to19}.
In this section, we consider four cases for link budget analysis, namely, space-ground, space-sea, ground-air-sea, and sea-air-sea links, and then the capacity of the link with various carrier frequencies is assessed and compared.
The band achieving the highest capacity is the most suitable frequency to support THz SAGIN, which balances the bandwidth and the molecular absorption loss.
Moreover, for these four scenarios, we contrast THz links of different structures, namely, with or without a plane at the altitude of 10~km as a relay node.
The antenna gain at each transceiver is set to be 55~dBi and the total transmit power is 10 W~\cite{Sen2023,9541155}.
In the case with a relay node as shown in~Fig.~\ref{fig_relay}, the total transmit power is allocated in the ratio that the two segments of the link achieve the same capacity.
Moreover, the water vapor distribution in the atmosphere is supposed to follow~\eqref{eq:nws} in saturated condition and T(0)=298.15~K.

\subsection{THz Space-Ground Link}
We first consider the wireless communication link between a satellite at an altitude of 550~km and a ground station as shown in Fig.~\ref{fig_relay}(a).
Furthermore, there are rain with a rain rate of 50 mm/hr and cumulonimbus with height of 5 km above the ground station.
Hence, the link La1 in Fig.~\ref{fig_relay}(a) is established across the rain and the cumulonimbus while La2 avoids crossing the cumulonimbus by setting the elevation angle to be $\pi/4$.
To avoid severe molecular absorption loss, the THz waves in 130-134 GHz, 167-174.8 GHz, and 209-226 GHz are adopted for the space-ground uplink, while that in 123-130 GHz, 130-134 GHz, and 167-174.8 GHz are adopted for the space-ground downlink.

The simulation results on THz space-ground uplink are given in TABLE~\ref{tab:spgup}, where $P_T$ and $P_R$ represent the transmit power and the received power, respectively. 
$P_N=N_0B$ is the noise power, where $N_0=-174~\textrm{dBm/Hz}$ denotes the power spectral density of the additive white Gaussian noise and $B$ represents the bandwidth. 
$C=B~\textrm{log}_2(1+P_R/P_N)$ is the capacity of the link.
The received power $P_R$ is acquired through link budget analysis and expressed as $P_R=P_T-L_{tot}+G$ in dB, where $G$ is the sum of the transmit gain and the receiver gain.
Moreover, $L_M$, $L_R$, and $L_{MO}$ refer to the loss caused by Mie absorption and scattering, the loss caused by Rayleigh absorption and scattering, and the loss caused by molecular absorption and scattering, respectively.
$L_M$, $L_R$, $L_{MO}$, and $L_{tot}$ are computed through the equations from~\eqref{eq:LC} to~\eqref{eq:nwu}.
For the space-ground uplink without relay nodes, the capacity of the THz link in 209-226~GHz is 0.01~Gbps while it is 34.35~Gbps if there are no rain and cumulonimbus.
Therefore, it is unrealistic to establish a THz space-ground uplink in rainy weather without relay nodes.
When there is a plane serving as a relay node, the space-ground uplink consists of an upper link between the spaceship and the plane and a lower link between the plane and the ground station.
The space-ground uplink utilizing 130-134 GHz in the lower link can provide higher capacity than that utilizing 167-174.8 GHz due to the fewer molecular absorption loss.
Moreover, the relay node can enable the space-ground uplink in 130-134 GHz and 209-226 GHz to achieve capacity of 45.41 Gbps, which is higher than the space-ground uplink crossing no rain or cumulonimbus without relay nodes.
Furthermore, the combination of La2 and Lb2 cannot provide higher capacity than the combination of La1 and Lb1 even the former scheme avoid the loss caused by the cumulonimbus, because the smaller elevation angle induces longer propagation distance in rain and humid atmosphere.

The simulation results on THz space-ground downlink are given in TABLE~\ref{tab:spgdown}.
For the space-ground downlink without relay nodes, the capacity of the THz link in 123-130~GHz is 4.62~Gbps while it is 67.77~Gbps if there are no rain and cumulonimbus.
For the space-ground downlink with a plane serving as a relay node, as in the space-ground uplink, the space-ground downlink utilizing 130-134 GHz in the lower link can provide higher capacity than that utilizing 167-174.8 GHz.
The relay node enables the space-ground downlink in 130-134 GHz and 123-130 GHz to achieve capacity of 45.34 Gbps, which is lower than the space-ground downlink crossing no rain or cumulonimbus without relay nodes due to smaller bandwidth of 130-134 GHz.
Hence, the THz space-ground downlink should not incorporate an aircraft as a relay node in sunny weather.
Due to the same reason in uplink, the combination of La2 and Lb2 cannot provide higher capacity than the combination of La1 and Lb1 for THz space-ground downlink.

In both THz space-ground uplink and downlink with a relay node, most power is allocated for the lower link in the troposphere. Hence, the capacity of the link in the troposphere is the bottleneck of THz SAGIN in space-ground communication scenario.

\begin{table*}[t]
\centering
\caption{THz space-sea uplink}
\label{tab:spsup}
\resizebox{\textwidth}{!}{%
\begin{tabular}{cccccccccccc}
Link & $f$ (GHz) & $P_T$ (W) & $L_{FP}$ (dB) & $L_{M}$ (dB) & $L_R$ (dB) & $L_{MO}$ (dB) & $L_{tot}$ (dB) & $P_R$ (dBm) & $P_N$ (dBm) & $C$ (Gbps) & $C/B$ (bps/Hz) \\ \hline
NRS & 252-265 & 10 & 195.50 & 0 & 0 & 28.15 & 223.65 & -73.65 & -72.86 & 11.37 & 0.87 \\ \hline
NRR & 252-265 & 10 & 195.50 & 21.04 & 29.44 & 28.15 & 274.13 & -124.13 & -72.86 & 0.00 & 0.00 \\ \hline
La3 & 130-134 & 9.85 & 154.86 & 22.70 & 12.53 & 6.67 & 196.76 & -46.83 & -77.98 & \multirow{2}{*}{40.98} & 10.35 \\
Lb3 & 252-265 & 0.15 & 195.34 & 0 & 0 & 0.31 & 195.65 & -63.89 & -72.86 &  & 3.15 \\ \hline
La3 & 167-174.8 & 9.96 & 157.10 & 22.05 & 18.03 & 20.37 & 217.55 & -67.57 & -75.08 & \multirow{2}{*}{21.24} & 2.73 \\
Lb3 & 252-265 & 0.04 & 195.34 & 0 & 0 & 0.31 & 195.65 & -69.63 & -72.86 &  & 1.63
\end{tabular}%
}
\end{table*}

\begin{table*}[t]
\centering
\caption{THz space-sea downlink}
\label{tab:spsdown}
\resizebox{\textwidth}{!}{%
\begin{tabular}{cccccccccccc}
Link & $f$ (GHz) & $P_T$ (W) & $L_{FP}$ (dB) & $L_{M}$ (dB) & $L_R$ (dB) & $L_{MO}$ (dB) & $L_{tot}$ (dB) & $P_R$ (dBm) & $P_N$ (dBm) & $C$ (Gbps) & $C/B$   (bps/Hz) \\ \hline
NRS & 123-130 & 10 & 189.30 & 0 & 0 & 7.11 & 196.41 & -46.41 & -75.55 & 67.77 & 9.68 \\ \hline
NRR & 123-130 & 10 & 189.30 & 22.76 & 11.74 & 7.11 & 230.91 & -80.91 & -75.55 & 2.58 & 0.37 \\ \hline
La3 & 130-134 & 9.85 & 154.86 & 22.70 & 12.53 & 6.67 & 196.76 & -46.83 & -77.98 & \multirow{2}{*}{41.40} & 10.35 \\
Lb3 & 123-130 & 0.15 & 189.14 & 0 & 0 & 0.30 & 189.44 & -57.68 & -75.55 &  & 5.96 \\ \hline
La3 & 167-174.8 & 9.98 & 157.10 & 22.05 & 18.03 & 20.37 & 217.55 & -67.56 & -75.08 & \multirow{2}{*}{21.32} & 2.73 \\
Lb3 & 123-130 & 0.02 & 189.14 & 0 & 0 & 0.30 & 189.44 & -66.43 & -75.55 &  & 3.20
\end{tabular}%
}
\end{table*}

\subsection{THz Space-Sea link}
We consider the wireless communication link between a spaceship at an elevation of 550~km and a ship as shown in Fig.~\ref{fig_relay}(b).
Moreover, there are rain with rain rate of 50 mm/hr and cumulonimbus with height of 5 km above the ship.
To balance the bandwidth and the propagation loss, the THz waves in 130-134 GHz, 167-174.8 GHz, and 252-265 GHz are adopted for the space-sea uplink, while that in 123-130 GHz, 130-134 GHz, and 167-174.8 GHz are adopted for the space-sea downlink.

The simulation results on THz space-sea uplink are given in TABLE~\ref{tab:spsup}.
For the space-sea uplink without relay nodes, the capacity of the THz link in 252-265~GHz is 0.00~Gbps while it is 11.37~Gbps if there are no rain and cumulonimbus.
Therefore, without relay nodes, THz space-sea uplink cannot be established in heavy rain.
When there is a plane serving as a relay node, the space-sea uplink utilizing 130-134 GHz in the lower link can provide higher capacity than that utilizing 167-174.8 GHz due to the fewer molecular absorption loss.
Moreover, the relay node can enable the space-sea uplink in 130-134 GHz and 252-265 GHz to achieve capacity of 40.98 Gbps, which is higher than the space-sea uplink without relay nodes in sunny weather.

The simulation results on THz space-sea downlink are listed in TABLE~\ref{tab:spgdown}.
For the space-sea downlink without relay nodes, the capacity of the THz link in 123-130~GHz is 2.58~Gbps while it is 67.77~Gbps in sunny weather.
For the space-sea downlink with a plane serving as a relay node, as in the uplink case, the space-sea downlink utilizing 130-134 GHz in the lower link can provide higher capacity than that utilizing 167-174.8 GHz.
The relay node enables the space-sea downlink in 130-134 GHz and 123-130 GHz to achieve capacity of 41.4 Gbps, which is almost twice of that in 167-174.8 GHz and 123-130 GHz.
However, it is still lower than the space-sea downlink without relay nodes in sunny weather due to the limited bandwidth of 130-134 GHz.
Hence, the THz space-sea downlink should not adopt an aircraft as a relay node in sunny weather.

For both THz space-sea uplink and downlink with a relay node, most power is allocated for the lower link in the troposphere to compensate the propagation loss. 
Hence, the capacity of the link in the troposphere is the bottleneck of THz SAGIN in space-sea communication scenario as in the space-ground communication scenario.

\begin{table*}[t]
\centering
\caption{THz ground-air-sea link}
\label{tab:gs}
\begin{threeparttable}
\resizebox{\textwidth}{!}{%
\begin{tabular}{cccccccccccc}
Link & $f$ (GHz) & $P_T$ (W) & $L_{FP}$ (dB) & $L_{M}$ (dB) & $L_R$ (dB) & $L_{MO}$ (dB) & $L_{tot}$ & $P_R$ (dBm) & $P_N$ (dBm) & $C$ (Gbps) & $C/B$ (bps/Hz) \\ \hline
NR & 141-148.5 & 10 & 175.66 & 0 & 0 & 314.6 & 490.26 & -340.26 & -75.25 & 0.00 & 0.00 \\ \hline
NR & 151.5-164 & 10 & 176.41 & 0 & 0 & 434.575 & 610.99 & -460.99 & -73.03 & 0.00 & 0.00 \\ \hline
NR & 209-226 & 10 & 179.20 & 0 & 0 & 777.005 & 956.20 & -806.20 & -71.70 & 0.00 & 0.00 \\ \hline
La4 & 141-148.5 & 5.00 & 169.81 & 0 & 0 & 41.95 & 211.76 & -64.77 & -75.25 & \multirow{2}{*}{27.04} & 3.60 \\
Lb4 & 141-148.5 & 5.00 & 169.81 & 0 & 0 & 41.95 & 211.76 & -64.77 & -75.25 &  & 3.60 \\ \hline
La4 & 151.5-164 & 5.00 & 170.56 & 0 & 0 & 56.12 & 226.67 & -79.68 & -73.03 & \multirow{2}{*}{3.53} & 0.28 \\
Lb4 & 151.5-164 & 5.00 & 170.56 & 0 & 0 & 56.12 & 226.67 & -79.68 & -73.03 &  & 0.28
\end{tabular}%
}
\begin{tablenotes}
    \footnotesize  
    \item[] NR denotes the link without relay nodes.
\end{tablenotes}
\end{threeparttable}
\end{table*}

\begin{table*}[t]
\centering
\caption{THz sea-air-sea link}
\label{tab:ss}
\resizebox{\textwidth}{!}{%
\begin{tabular}{cccccccccccc}
Link & $f$ (GHz) & $P_T$ (W) & $L_{FP}$ (dB) & $L_{M}$ (dB) & $L_R$ (dB) & $L_{MO}$ (dB) & $L_{tot}$ & $P_R$ (dBm) & $P_N$ (dBm) & $C$ (Gbps) & $C/B$ (bps/Hz) \\ \hline
NR & 141-148.5 & 10 & 167.70 & 0 & 0 & 125.84 & 293.54 & -143.54 & -75.25 & 0.00 & 0.00 \\ \hline
NR & 151.5-164 & 10 & 168.45 & 0 & 0 & 173.83 & 342.28 & -192.28 & -73.03 & 0.00 & 0.00 \\ \hline
NR & 209-226 & 10 & 171.40 & 0 & 0 & 310.80 & 482.20 & -332.20 & -71.70 & 0.00 & 0.00 \\ \hline
La5 & 141-148.5 & 5.00 & 162.65 & 0 & 0 & 18.40 & 181.05 & -34.06 & -75.25 & \multirow{2}{*}{102.62} & 13.68 \\
Lb5 & 141-148.5 & 5.00 & 162.65 & 0 & 0 & 18.40 & 181.05 & -34.06 & -75.25 &  & 13.68 \\ \hline
La5 & 151.5-164 & 5.00 & 163.40 & 0 & 0 & 24.61 & 188.01 & -41.02 & -73.03 & \multirow{2}{*}{132.93} & 10.63 \\
Lb5 & 151.5-164 & 5.00 & 163.40 & 0 & 0 & 24.61 & 188.01 & -41.02 & -73.03 &  & 10.63 \\ \hline
La5 & 209-226 & 5.00 & 166.19 & 0 & 0 & 45.25 & 211.44 & -64.45 & -71.70 & \multirow{2}{*}{45.15} & 2.66 \\
Lb5 & 209-226 & 5.00 & 166.19 & 0 & 0 & 45.25 & 211.44 & -64.45 & -71.70 &  & 2.66
\end{tabular}%
}
\end{table*}

\subsection{THz Ground-Air-Sea link}
We consider the wireless communication link between a ground station and a ship in a distance of 100~km from the ground station as shown in Fig.~\ref{fig_relay}(c).
Moreover, the weather above the ground station and the ship is supposed to be sunny.
To fully assess the capacity of THz ground-sea link, the THz waves in 141-148.5 GHz, 151.5-164 GHz, and 209-226~GHz are adopted for the ground-sea communication scenario.
As shown in TABLE~\ref{tab:spsup}, the THz ground-air-sea link cannot be established without relay nodes due to the enormous molecular absorption loss in the propagation distance of 100~km.
For THz ground-air-sea link with a relay node, only the band from 141 to 148.5~GHz can enable communication and achieve capacity of 27.04~Gbps.
Hence, THz SAGIN should utilize the band 141-148.5~GHz in ground-sea communication scenario.
However, if the communication distance further increases, THz ground-air-sea link cannot be established with only one single relay node.
In such situation, higher antenna gain or more relay nodes are necessary.
\subsection{THz Sea-Air-Sea Link}
We consider the wireless communication link between two ships and they are in the distance of 40~km from each other as shown in Fig.~\ref{fig_relay}(d).
Moreover, the weather above the sea is supposed to be sunny.
To fully assess the capacity of THz sea-air-sea link, the THz waves in 141-148.5 GHz, 151.5-164 GHz, and 209-226~GHz are adopted for the sea-sea communication scenario.

The simulation results on THz sea-air-sea link are listed in TABLE~\ref{tab:spsup}.
THz sea-sea link cannot be established without relay nodes since the link has to propagate in the humid air which induce enormous molecular absorption loss.
When there is a plane serving as a relay node, the sea-sea link in 151.5-164 GHz achieves highest capacity of 132.93~Gbps due to the balance between bandwidth and propagation loss, while it is 102.62~Gbps for 141-148.5~GHz, and 45.15~Gbps for 209-226~GHz.
Hence, the band in 151.5-164~GHz achieves the highest capacity in the sea-sea communication scenario in THz SAGIN.

\section{Conclusion}\label{sec:conclusion}
In this paper, we proposed a universal attenuation model for the THz SAGIN, where the absorption and scattering effects induced by condensed particles, molecules, and free electrons are respectively unified, ultimately presenting a physical picture of THz waves.
Specifically, the collision model between photons and medium particles is developed, where the absorption and scattering cross-sections are used to characterize the intensity of collision. 
Based on this model, the total propagation loss and the link capacity of THz space-ground, space-sea, ground-sea, and sea-sea scenarios are evaluated.
Some take-away lessons about the capacity and network design for THz SAGIN are summarized as follows.
First, 
numerical results of the achievable rates for different frequency bands demonstrate that the 130-134~GHz band can achieve the highest capacity in space-sea and space-ground scenarios while the 151.5-164~GHz band achieves the highest capacity in sea-sea scenario at distance lower than approximately 40~km.
Second, due to the abundant medium particles including water vapor, raindrops and oxygen, the wireless link in the troposphere is the bottleneck of the entire THz space-air-ground link.
Third, the relaying strategy cannot increase the capacity of 
the THz downlink channel for space-sea and space-ground scenarios in sunny days.
Compared with ITU standards, our proposed physical picture of THz wave provides an interpretable attenuation model for the THz SAGIN, which could be a good compensation for current standards on atmospheric attenuation.



%


\bibliographystyle{IEEEtran}
\bibliography{IEEEabrv,reference}

\end{document}